\documentclass[a4paper,12pt]{article}
\usepackage{exscale}
\usepackage{amssymb,amsmath,latexsym,theorem}
\def \slas{\kern -6.2pt /}
\def \sla{\kern -5.4pt /}
\def \sl{\kern -4.0pt /}
\def \Cslas{\kern -6.8pt /}
\def \Dslas{\kern -7.4pt /}
\def \slass{\kern -7.4pt /}
\def \ii{\mathrm{i}}
\def \d{\mbox{d}}

\def \pd{\partial}

\def \lcxg{x\slas}
\def \lcyg{y\slas}
\def \lcx{\tilde{x}}
\def \tl#1{\overset{\kern 2pt\circ}{#1}}
\def \TL#1{\overset{\kern -28pt \circ}{#1}}
\def \TLL#1{\overset{\kern -7pt \circ}{#1}}
\def \relstack#1#2{\mathrel{\mathop{#2}\limits_{#1}}}
\renewcommand{\theequation}{\arabic{section}.\arabic{equation}}
\renewcommand{\[}{\left[}
\renewcommand{\]}{\right]}

\renewcommand{\phi}{\varphi}
\renewcommand{\theta}{\vartheta}

\newcommand{\bq}{\begin{equation}}
\newcommand{\eq}{\end{equation}}
\newcommand{\bea}{\begin{eqnarray}}
\newcommand{\eea}{\end{eqnarray}}

\newcommand\ka{\kappa_1}
\newcommand\kb{\kappa_2}

\newcommand\xx{\tilde{x}}

\begin{document}
%

%%%%%%%%%%%%%%%%%%%%%%%%%%%
%%  Datei: laztitle.tex  %%
%%%%%%%%%%%%%%%%%%%%%%%%%%%

\title{\bf Decomposition of nonlocal light-cone operators into
     harmonic operators of definite twist}

\author{\large Bodo~Geyer\thanks{
e-mail: geyer@itp.physik.uni-leipzig.de}\\
       % \emph{}\\
        \emph{Universit\"at~Leipzig, Institut~f\"ur~Theoretische~Physik}\\
        \emph{and Center for Theoretical Studies}\\
        \emph{Augustusplatz~10, D-04109~Leipzig, Germany}\\[2ex]
        %\emph{}\\
        %\large 
Markus~Lazar\thanks{e-mail: lazar@itp.physik.uni-leipzig.de}
\\
        %\emph{Universit\"at~Leipzig}\\
        \emph{Universit\"at~Leipzig, Institut~f\"ur~Theoretische~Physik}\\
        %\emph{Augustusplatz~10}\\
        \emph{Augustusplatz~10, D-04109~Leipzig, Germany}\\[2ex]
     \large{Dieter~Robaschik}\thanks{
present address: DESY-IfH Zeuthen, Platanenallee 6, 15735 Zeuthen}\\    
     \emph{Karl--Franzens--Universit\"at~Graz,
     Institut~f\"ur~Theoretische~Physik}\\
        %\emph{}\\
        %\emph{}\\
        \emph{Universit\"atsplatz~5, A-8010~Graz, Austria}}
\date{}
\maketitle
\thispagestyle{empty}
%
%
%\vspace{1cm}
%
%\vfill
%\newpage
%\thispagestyle{empty}
\begin{abstract}
\noindent
Bilocal light--ray operators which are Lorentz scalars, 
vectors or antisymmetric tensors, and 
which appear in various hard QCD scattering
processes, are decomposed into operators of definite 
twist. These operators are harmonic tensor functions and 
their Taylor expansion consists of (traceless) 
local light--cone operators which span irreducible
representations of the Lorentz group with definite spin $j$
and common (geometric) twist (= dimension -- spin). 
Some applications concerning the nonforward matrix elements
of these operators and the generalization to conformal
light--cone operators of definite twist is considered.
The group theoretical background of the method has been made clear.
\end{abstract}

%\vspace{1cm}
%\begin{flushleft}
\noindent
NTZ 2/99 \hspace{.5cm} UNIGRAZ-UTP/27-01-99 
\hfill {\tt hep-th/9901090}\\
%27-01-99


\newpage

%%%%%%%%%%%%%%%%%%%%%%%%%%%
%%  Datei: lazintro.tex  %%
%%%%%%%%%%%%%%%%%%%%%%%%%%%

%%%%%%%%%%%%%%%%%%%%%%%%%%%%%%%%%%%%%%%%%%%%
%%%%%%%%%%%%%%%%%%%%%%%%%%%%%%%%%%%%%%%%%%%%
\section{Introduction}
\renewcommand{\theequation}{\thesection.\arabic{equation}}
\setcounter{equation}{0} 
%\setcounter{ footnote}{0}
%%%%%%%%%%%%%%%%%%%%%%%%%%%%%%%%%%%%%%%%%%%%
%%%%%%%%%%%%%%%%%%%%%%%%%%%%%%%%%%%%%%%%%%%%
%
For many hard scattering processes in QCD the 
nonlocal light--cone(LC) expansion, together with the 
renormalization group equation, is a powerful tool to 
determine the dependence of the nonperturbative 
partition amplitudes on the experimentally relevant
momentum transfer $Q^2$. This is the case for,
e.g.,~the parton distributions in deep inelastic
scattering (DIS), the pion as well as vector meson wave 
functions and, as growing up more recently, the
non--forward distribution amplitudes in deeply virtual
Compton scattering (DVCS). Any of these phenomenological 
quantities are related to the Compton amplitude for non--forward
scattering a virtual photon off a hadron which is given by
%-------------------------------------------------------
\begin{equation}
\label{COMP}
T_{\mu\nu}(p_+,p_-,Q) 
= i \int d^4x \,e^{iqx}\,
\langle p_2, S_2\,|T (J_{\mu}(x/2) J_{\nu}(-x/2))|\,p_1,
S_1\rangle,
\end{equation}
%-------------------------------------------------------
where $p_+ = p_2 + p_1,\; p_- = p_2 - p_1 = q_1 - q_2, \;
q = (q_1 + q_2)/2$ are the kinematical independent variables.
To specify the asymptotics of 
virtual Compton scattering the generalized Bjorken 
region will be defined as follows
\begin{eqnarray}
\label{gBr1}
\nu =  qp_+ \longrightarrow \infty, \qquad
Q^2 = - q^2 \longrightarrow \infty
\end{eqnarray}
with the two independent scaling variables
%\begin{eqnarray}
%\label{gBr2}
$\xi  =  {Q^2 }/{qp_+}$ and
$\xi' = {qp_-}/{qp_+} = {(q_1^2 - q_2^2)}/{2\nu},
$ 
to be held fixed by the experimental setup. These variables
which are not restricted to the intervall $[-1, +1]$ are
obvious modifications of the usual Bjorken variable
and the 'skewedness' parameter.\footnote{
%%%%%%%%%%%%%%%%%%%%%%%%%%%%%%%%%%%%%%%%%
Instead of this parametrization, despite being quite natural 
from a general point of view,
for various physical situations others may be prefered
(see, e.g.,~\cite{XJ,RAD,Jaf96,Gui98}).
}
%%%%%%%%%%%%%%%%%%%%%%%%%%%%%%%%%%%%%

The (renormalized) time--ordered product in eq.~(\ref{COMP})
can be represented in terms of the {\em nonlocal} operator  
product expansion~\cite{AS78, BNL, LEIP}, which directly leads
to compact expressions for the coefficient functions and
the corresponding bilocal LC--operators. 
It is essential to remark that all the interesting 
nonperturbative partition amplitudes which appear in the
above mentioned processes are Fourier transforms of 
various matrix elements whose unique input are
one and the same light--ray operators. 
Furthermore, all the interesting evolution equations of 
these partition amplitudes result from the renormalization group
equation of the involved light--ray operators and,
consequently, their evolution kernels can be determined from the 
anomalous dimension of these operators. So many of the 
physical properties which result in different experimental  situations 
have a common quantum field theoretical origin which is traced back
to the properties of the LC--operators.

With the growing accuracy of the experimental data a
perturbative expansion with respect to the 
experimentally relevant variable, $M^2/Q^2$, $M$ being a 
typical mass of the process, will be very useful. However, 
these contributions  could be of quite different origin. On the
one hand, they result from a decomposition of the bilocal 
light--ray operators -- in lowest order only quark operators 
with two external legs occure --  into operators of definite 
twist.  In addition there appear higher loop contributions of 
physical matrix elements for the same operators. 

Unfortunately, the notion of 
twist is unambigously defined in leading order only.
Its original definition, as has been introduced by Gross and
Treiman \cite{Gro71} for the {\em local} operators, 
twist $(\tau)$ = dimension $(d)$ -- spin $(j)$, is directly 
related to the irreducible tensor representations of the orthochronous
Lorentz group and, with respect to this, is a Lorentz invariant 
notion ("geometric twist"). 

Later on, various authors used the
well known LC--quantization in the infinite momentum frame
\cite{KS70} together with the decomposition of the quark fields into
"good" and "bad" components, $\psi = \psi_+ + \psi_-$, with
$\psi_\pm = \frac{1}{2}\gamma^\mp\gamma^\pm \psi$, in order 
to introduce another notion of twist: As has been
pointed out in \cite{Jaf92}, a "bad" component introduces one
unit of $M/Q$ because these components are not
dynamically independent and may be expressed through the 
equations of motion by the "good" ones times the above
 kinematical factor. Despite being 
conceptionally different this definition looks similar to the
phenomenologically more convenient one which counts only 
powers of $1/Q$ in the infinite momentum frame \cite{Jaf92}.

However, both these power--counting concepts 
of "dynamical twist", despite being phenomenological quite useful,
have serious theoretical drewbacks: First, they are not Lorentz 
invariant and, therefore, can not coincide with the original
geometric definition. Furthermore, the same power of $1/Q$ may 
occure for different values of $\tau$. In addition, this 
concept is {\em applicable only for matrix elements} of operators
 and not directly related to the operators itself. 
Moreover, if the matrix elements contain more than
one momentum, as will be the case for DVCS, the definition
of the scaling variables is somewhat ambiguous and different 
definitions are related by power series in $M^2/Q^2$
thereby changing the decomposition with respect to 
"dynamical twist".

Therefore, from the point of view of renormalized quantum
field theory we should state: Only the original geometric 
definition of twist is a well-defined concept. But then the 
problem occurs how to generalize it to the nonlocal operators
which appear in the nonlocal light--cone expansion.

Let us shortly introduce the basic concepts of the nonlocal
light--cone expansion and show how it is related to its local
version. The rigorous proof of the nonlocal LCE was given 
in \cite{AS78} 
where, first of all, an {\em operator identity}
for the product of two composite operators -- 
e.g.~the (time ordered) product of the two currents 
in eq.~(\ref{COMP}) -- has been proven 
(in scalar field theory)
which holds on the whole Hilbert space.
Then, using a special, light--cone adapted 
renormalization procedure
$R$ it was shown that the perturbative functional 
of the renormalized product of two currents may be
split into an asymptotically relevant part -- the light--cone
expansion  up to a finite order of light--cone singularity  --
and a well--defined remainder being less singular:
\begin{eqnarray}
\label{GLCE}
\hspace{-.5cm}
{R} T\big(J_\mu(x/2)J_\nu(-x/2)S\big) 
&\approx&
\int^{+1}_{-1} \!\!d^2 \underline\kappa\,
C_\Gamma(x^2, \underline\kappa; \mu^2) 
{R }T\!\left( O^\Gamma(\ka \xx, \kb \xx) S\right)
\\
\hspace{-.5cm}
&&\qquad+\;{\rm higher~order~terms.}
\nonumber
\end{eqnarray}
Here, the (unrenormalized) light--ray operators 
 are given by
\begin{eqnarray}
\label{GLCEOP}
O^\Gamma(\ka \xx, \kb \xx) 
=
\;:\overline\psi(\ka\xx) \Gamma 
U(\ka\xx, \kb\xx) \psi(\kb\xx):\ ,
\end{eqnarray}
with some specified $\Gamma$--structure 
and with the usual phase factor
\begin{eqnarray}
\label{phase}
U(\ka, \kb)
\equiv
U(\ka \xx, \kb \xx) 
 = 
{\cal P} \exp\left\{-ig
\int^{\ka}_{\kb} d\tau \,\xx^\mu A_\mu (\tau \xx)
\right\},
%\nonumber
\end{eqnarray}
where ${\cal P}$ denotes the path ordering, 
$g$ is the strong coupling parameter 
and $ A_\mu = A_\mu^a t^a$ is the gluon field with
$t^a$ being the generators of $SU(3)_{\rm color}$
in the fundamental representation spanned 
by the quark fields $\psi$.
Thereby, the light--like vector
%-------------------------------------------------------------
\begin{equation}
\label{LCV}
\xx = x + \eta\, {(x\eta}/{\eta^2)}
\left(
\sqrt{1 - {x^2 \, \eta^2}/{(x\eta)^2}} - 1 
\right)
\end{equation}
%-------------------------------------------------------------
is related to $x$ and a fixed non--null subsidiary four--vector
$\eta$ whose dependence drops out in physical 
expressions at leading order.\footnote{
It has to be noted that, because the light--cone expansion is
considered for the operators and {\em not} for their {\em matrix
elements}, such a light--like reference vector has necessarily 
to be introduced in configuration space. However, if matrix
elements are considered with some uniquely defined momentum
$P, P^2 = M^2,$ then $\eta = P/M$ may be chosen \cite{BBKT98}.
} 
In this approach the coefficient functions and the corresponding 
light--ray operators naturally occure as renormalized ones. 
In the following, for simplicity, we do not explicitly indicate 
that they are renormalized quantities.

Let us now point to the remarkable fact that, in principle,
the nonlocal as well as the usual local light--cone 
expansion may be obtained from (\ref{GLCE})
(for a detailed consideration see \cite{BGHR85,SLAC}):
The Fourier transform $F_\Gamma(x^2, \xx p_i; \mu^2)$ 
of $C_\Gamma(x^2, \kappa_i; \mu^2)$ with respect to 
$\kappa_i$ is an entire functions of the new arguments 
$\xx p_i$ \cite{AS78}. Therefore, the range of 
$\kappa_i$ is restricted to $-1 \leq \kappa_i \leq +1$
and, in addition, $F_\Gamma$ may be Taylor expanded with respect to 
$\xx p_i$: this leads, instead of the $\kappa$--integration
in (\ref{GLCE}), to a (twofold) infinite 
sum over local operators. The local and nonlocal
operators are related through the following formulae:
\begin{eqnarray}
\label{taylor}
O^\Gamma(\ka \xx, \kb \xx) 
&=&
\sum_{n_1 n_2}
\frac{\ka^{n_1}}{n_1!}
\frac{\kb^{n_2}}{n_2!} \;
O^\Gamma_{n_1 n_2}(\xx), 
\\
\label{dtaylor}
O^\Gamma_{n_1 n_2}(\xx)
&=&
\frac{\partial^{n_1}}{\partial \ka^{n_1}}
\frac{\partial^{n_2}}{\partial \kb^{n_2}}
\left. O^\Gamma(\ka \xx, \kb \xx) 
\right|_{\ka=\kb=0}.
\end{eqnarray}

This connection between the local and the nonlocal LC--operators
is the key relation which opens the possibility to generalize 
the geometric notion of twist, being originally introduced for local 
operators, also to nonlocal ones. This is the aim of the
present paper which is essentially based on \cite{LAZAR,BGR98}.
At first, a group theoretical founded algorithm is presented
which allows to obtain for any bilocal operator
a uniquely defined  twist decomposition
\bea
O_\Gamma(\ka x, \kb x) 
= 
\sum_{\tau_{\rm min}}^\infty
O_\Gamma^\tau(\ka x, \kb x).
\eea
For general values of $x$ this decomposition contains infinitely
many terms which are harmonic tensor functions (depending
on the "external" Lorentz indices $\Gamma$). In general,
the contributions to a well-defined twist result from various
independent irreducible (local) tensor representations.
By projecting onto the light--cone, $x \rightarrow \xx$,
the above sum terminates at finite twist $\tau_{\rm max}$. For the
bilocal quark operators, which will be considered in detail,
we obtain $\tau_{\rm max}=4$. For the vector and (antisymmetric) 
tensor operators  the trace terms which define
operators of higher twist are proportional to $\xx_\beta$. Their
explicit form has been given here for the first time.
Of course, the method can be applied to the gluon operators; 
in principle, also more complicated nonlocal operators may be
considered.

It should be mentioned here that a special version of that programme 
making extensive use of Schwinger--Fock gauge for the 
quark string operators, however without unravelling the 
explicit symmetry type behind such twist decompositions, already 
was given in Ref.~\cite{BB88}. In addition we remark that a
twist decomposition of special conformal operators
was given in Ref.~\cite{D}~\footnote{
The authors are much indebted to D.~M\"uller for bringing
that referenc to their attention.} thereby taking profit from the 
properties of conformal invariant 3--point functions of
(generalized) free fields.

The paper is organized as follows. Section 2 is 
devoted to a short exposition of the group theoretical 
background  characterizing the irreducible tensor
representations of the Lorentz group. Section 3 contains our
main results -- the twist decomposition of (pseudo) scalar, 
(axial) vector and antisymmetric  tensor operators. 
%Section 4 gives the
%relation of the twist$(\tau)$ decomposition to the other
%decompositons for the case of DIS. 
Appendix \ref{young} contains a
short exposition about the characterization of irreducible
tensor representations through the Young tableaux', and 
Appendix \ref{trace} introduces the notion of harmonic tensor functions
which is relevant for our twist decomposition.

%%%%%%%%%%%%%%%%%%%%%%%%%%%%
%%%  Datei: lazgroup.tex %%%
%%%%%%%%%%%%%%%%%%%%%%%%%%%%

\section{Irreducible tensor representations of
the \\ Lorentz group
${\cal L}^\uparrow_+ \simeq SO(1,3,{\Bbb R})$}
\renewcommand{\theequation}{\thesection.\arabic{equation}}
\setcounter{equation}{0}

In this Chapter the group theoretical background will be given
which is used for the construction of the nonlocal operators
whose local parts -- which are obtained by Taylor expansion --
transform according to irreducible tensor representations
 under the Lorentz group,
i.e.,~have definite spin and therefore also well defined
twist. These tensor representations are characterized by a 
specific symmetry class with respect to the symmetric group
and are determined by a few types of Young tableaux.

The Lie algebra of the Lorentz group ${\cal L}^\uparrow_+ $ 
is characterized by the generators of the three spatial
rotations $\vec M$ and the three boosts $\vec N$; from it the 
two (complex) linear combinations ${\vec M}_\pm$ may be build
which define two independent $SO(3)$ groups:
\begin{eqnarray}
\[M_\pm^i, M_\pm^j \] ~=~i \epsilon^{ijk} M_\pm^k,
\qquad
\[ M_+^i, M_-^j \] ~=~ 0
\quad {\rm with} \quad
{\vec M}_\pm = {\vec M} \pm i {\vec N}.
\nonumber
\end{eqnarray}
Therefore, the (Lie algebra of the) complex Lorentz group
is isomorphic to the (Lie algebra of the) direct product
$SO(3, {\Bbb C}) \otimes SO(3,{\Bbb C)} \simeq SO(4,{\Bbb C})$.
This characterization makes use of the fact that the irreducible 
(finite) representations 
\begin{eqnarray}
{\cal D}^{(j_+,j_-)}({\vec \phi}, {\vec \theta})
~=~ 
{\cal D}^{(j_+)}({\vec \phi} - i {\vec \theta}) 
\otimes
{\cal D}^{(j_-)}({\vec \phi} + i {\vec \theta})
\end{eqnarray} 
of the restricted orthochronous Lorentz group 
${\cal L}^\uparrow_+ \simeq SO(1,3;{\Bbb R})$, 
where $\vec \phi$ and $\vec \theta$ are the angle of rotation 
and the rapidity of the boost transformations, 
are determined by two numbers 
$(j_+, j_-), j_\pm = 0,1/2, 1, \cdots $, which define the
spin $j = j_+ + j_-$ of the representation.

Let us now state some general results concerning finite
dimensional representations of the (complex)
orthogonal groups (see, e.g.,~Chapters 8 and 10 of
Ref.~\cite{BR}):

%\begin{itemize}
%\item
\noindent $(1)~~$
The group $SO(N, {\Bbb C})$ has two series of 
complex--analytic\footnote{
A representation is complex--analytic if it depends
analytically on the group parameters.}
irreducible representations. Every representation of the first
(resp. the second) series determines and is in turn determined 
by a highest weight ${\underline m} = (m_1, m_2, \ldots ,
m_\nu)$ 
whose components $m_i$ are
integers (resp. half--odd integers) and satisfy the conditions
\begin{eqnarray}
\label{even}
m_1 \geq m_2 \geq \ldots \geq m_{\nu -1} \geq|m_\nu|
&{\rm for}& N = 2\nu,
%\nonumber
\\
\label{odd}
m_1 \geq m_2 \geq \ldots \geq m_{\nu -1} \geq m_\nu \geq 0
&{\rm for}& N = 2\nu +1.
%\nonumber
\end{eqnarray}
The first series determines the tensor representations,
 whereas the second series determines the spinor
representations.

%\item
\noindent$(2)~~$
A tensor representation of $SO(N; {\Bbb C})$, for either
$N = 2\nu$ or $N = 2\nu +1$, being determined
by the highest weight ${\underline m} = 
(m_1, m_2, \ldots , m_\nu), \;m_i$ integer, is 
equivalent to another tensor representation
$T_{i_1 i_2 \ldots i_n}, \; n = \sum m_i,$ which is
realized in the space of traceless tensors whose 
symmetry class is characterized by a 
Young pattern defined by the partition 
$[m]=  (m_1, m_2, \ldots , m_\nu)$. 

%\item
\noindent$(3)~~$
Furthermore, if ${\cal T}_{G_c}$ is a complex--analytic
representation of a (complex) semi-simple Lie group $G_c$
and ${\cal T}_{G_r}$ is the restriction of ${\cal T}_{G_c}$
to a real form $G_r$ of $G_c$, then, ${\cal T}_{G_c}$
is irreducible (fully reducible) iff ${\cal T}_{G_r}$ is
irreducible (fully reducible).
%\end{itemize}
%
As a consequence,
any irreducible representation of the complex group remains
irreducible if restricted to a real subgroup and, 
on the other hand,
from any irreducible representation of the real group by 
analytic continuation in the group parameters 
an irreducible representation of its complexification is
obtained.

Therefore, in order to investigate irreducible tensor 
representations of the Lorentz group we may consider equally
well irreducible tensor representations of the complex
group $SO(4,{\Bbb C})$. Even more, any irreducible 
representation of the orthogonal group $SO(4)$ by analytic 
continuation induces an irreducible representation of 
$SO(4,{\Bbb C})$, which by restriction to their real subgroup 
${\cal L}^\uparrow_+$ subduces
an irreducible representation of the Lorentz group. Since
tensor representations of the orthogonal group are uniquely
determined by the symmetry class $[m]$ of their tensors 
we may study the irreducible tensor representations of 
the Lorentz group through a study of the corresponding 
Young tableaux which determine the irreducible representations
of the symmetric group $S_n$ of permutations.

This characterization of tensor representations through their 
symmetry class holds for any of the classical matrix groups, 
$GL(N,{\Bbb C})$ and their various subgroups 
 (see also Appendix \ref{young}).
However, for $O(N,{\Bbb C})$ and $SO(N,{\Bbb C})$ these 
representations, in general, are not irreducible. 
The reason is that taking the trace of a tensor 
commutes with the orthogonal transformations.
Therefore, by Schur's Lemma, irreducible subspaces 
of $SO(N,{\Bbb C})$ are spanned by {\em traceless}
tensors having definite symmetry class.

From the requirement of tracelessness the following 
restrictions obtain (see, e.g.,~\cite{Ham62}):
The only Young tableaux' being relevant are those whose first
two columns are restricted to have lenght 
$\mu_1 + \mu_2 \leq N$ (see (i) -- (iv) below). 
Two representations $R$ and $R'$ 
whose first columns are related by $\mu'_1 = N - \mu_1$, 
where $\mu_1 \leq N/2$, are called associated; 
if $\mu_1 = \mu'_1 = N/2$ this representation
is called selfassociated. After restriction to the subgroup
$SO(N) \subset O(N)$ associated representations are equivalent,
whereas selfassociated representations decompose into two
nonequivalent irreducible representations. 

Let us illustrate this for the Lorentz group by the 
representations which will be of interest in the following.
First of all, vector and axial vector representations 
$V_\mu$ and $A_\mu = \epsilon_{\mu\nu\kappa\lambda}
A^{\nu\kappa\lambda}$, respectively, are associated ones, 
and antisymmetric tensor representations,
$A_{\mu\nu} = T_{\mu\nu} - T_{\nu\mu}$, are selfassociated
which, if restricted to $SO(1,3)$ decompose
into the (anti-)selfdual tensors
$A^\pm_{\mu\nu} = \frac{1}{2}( A_{\mu\nu} \mp \frac{1}{2} \, 
\ii \epsilon_{\mu\nu\kappa\lambda} A^{\kappa\lambda})$.
Since later on only representations of
the orthochronous Lorentz group containing the parity operation
are considered, this distinction is of no relevance.
Therefore, any tensor of second order, $T_{\mu\nu}$, may be 
decomposed according to
\begin{eqnarray}
\label{2tensor}
T_{\mu\nu}
&=&
\underbrace{
\hbox{\large $\frac{1}{2}$}
\left(T_{\mu\nu}+T_{\nu\mu}\right)
-
\hbox{\large $\frac{1}{4}$}
g_{\mu\nu} T^{\;\rho}_\rho 
     }_{S_{\mu\nu}} 
+ 
\underbrace{
\hbox{\large $\frac{1}{2}$}
\left(T_{\mu\nu}-T_{\nu\mu}\right) 
     }_{A_{\mu\nu}}
+
\hbox{\large $\frac{1}{4}$}
g_{\mu\nu} T^{\;\rho}_\rho .
\end{eqnarray}
Let us denote by ${\bf T}(j_+, j_-),\, j_+ + j_-$ integer, 
the space of tensors
which carry an irreducible representation ${\cal D}^{(j_+,j_-)}$
of the Lorentz group. Then from eq.~(\ref{2tensor}) we read off:
$S_{\mu\nu} \in {\bf T}(1,1)$ is a symmetric traceless tensor, 
$A_{\mu\nu} = A^+_{\mu\nu} + A^-_{\mu\nu} \in 
{\bf T}(1,0) \oplus {\bf T}(0,1)$ are the selfdual
and the antiselfdual antisymmetric tensors, and
$\hbox{$\frac{1}{4}$} g_{\mu\nu} T^{\;\rho}_\rho \in 
{\bf T}(0,0)$ corresponds to the trivial representation defined
through the unit tensor. This decomposition corresponds to the
Clebsch-Gordan decomposition of 
the direct product of two vector representations:
\begin{eqnarray}
\Big(
\hbox{\large $\frac{1}{2}$}, \hbox{\large $\frac{1}{2}$}
\Big)\otimes\Big(
\hbox{\large $\frac{1}{2}$}, \hbox{\large $\frac{1}{2}$}
\Big)
&=&
(1,1) \oplus\Big( (1,0)\oplus(0,1) \Big) \oplus (0,0) .
\end{eqnarray}

Now we consider the decomposition of the space of tensors of
rank $n$ into irreducible representation spaces of 
$SO(1,3; {\Bbb R})$. The different symmetry classes are strongly
restricted by the requirement that only such Young patterns
$[m]$
are allowed for which the sum of the first two colums is lower
or equal to four. Therefore, only the following Young patterns
correspond to nontrivial irreducible representations
by traceless tensors:
\newpage
\begin{enumerate}
\item[i.] \unitlength0.5cm
\begin{picture}(30,1)
\linethickness{0.15mm}
\multiput(1,0)(1,0){13}{\line(0,1){1}}
\put(1,1){\line(1,0){12}}
\put(1,0){\line(1,0){12}}
\put(15,0){$j=n,n-2,n-4,\ldots$}
\end{picture}
\item[ii.]\unitlength0.5cm
\begin{picture}(5,1)
\linethickness{0.15mm}
\multiput(3,0)(1,0){10}{\line(0,1){1}}
\multiput(1,-1)(1,0){2}{\line(0,1){2}}
\put(1,1){\line(1,0){11}}
\put(1,0){\line(1,0){11}}
\put(1,-1){\line(1,0){1}}
\put(15,0){$j=n-1,n-2,n-3,\ldots$}
\end{picture}
\item[iii.] \unitlength0.5cm
\begin{picture}(5,2)
\linethickness{0.15mm}
\multiput(3,0)(1,0){9}{\line(0,1){1}}
\multiput(1,-2)(1,0){2}{\line(0,1){3}}
\put(1,1){\line(1,0){10}}
\put(1,0){\line(1,0){10}}
\put(1,-1){\line(1,0){1}}
\put(1,-2){\line(1,0){1}}
\put(15,0){$j=n-2,n-3,\ldots$}
\end{picture}
\item[iv.] \unitlength0.5cm
\begin{picture}(5,3)
\linethickness{0.15mm}
\multiput(4,0)(1,0){8}{\line(0,1){1}}
\multiput(1,-1)(1,0){3}{\line(0,1){2}}
\put(1,1){\line(1,0){10}}
\put(1,0){\line(1,0){10}}
\put(1,-1){\line(1,0){2}}
\put(15,0){$j=n-2,n-3,\ldots$}
\end{picture}
\end{enumerate}
\vspace*{3mm}
In addition, for $n = 4$, also the completely antisymmetric 
tensor of rank 4 which is proportional to 
$\epsilon_{\mu\nu\kappa\lambda}$, and therefore equivalent to
the trivial representation is allowed. 

For the cases (i) -- (iv)
the minimal spin $j$ -- depending on $n$ being either even or
odd --
will be zero or one. In the case of symmetry type (iv) we have
given only one special Young pattern; in principle the lenght
of the second row  may contain up to 
$m_2 = [\frac{n}{2}]$ boxes, and then the maximal spin is
given by $j = n - m_2$. ---
The representations corresponding to symmetry class (i) are
associated to representations of the symmetry class (iii) with
$n+2$ boxes. The representations corresponding to symmetry class 
(ii) and (iv) are selfassociated. The symmetry class (ii) 
contains two non-equivalent parts being related to  
$(\frac{n}{2},\frac{n}{2}-1)$ and 
$(\frac{n}{2}-1,\frac{n}{2})$; the symmetry class (iv) 
contains three nonequivalent parts related to
$(\frac{n}{2},\frac{n}{2}-2)$, $(\frac{n}{2}-1,\frac{n}{2}-1)$ 
and $(\frac{n}{2}-2,\frac{n}{2})$; and so on.
Any tensor whose symmetry class does not coincide with one 
of the above classes vanishes identically due to the requirement
of tracelessness.

There are two possible ways to construct the nonvanishing 
tensors. Either one symmetrizes the indices according to the
corresponding (standard) Young tableaux and afterwards subtracts 
the traces, or one starts from tensors being already traceless 
and finally symmetrizes because this does not destroy the
tracelessness. For practical reasons the latter procedure seems
to be preferable and will be used in the construction of
irreducible light--cone operators of definite twist. 
%%%%%%%%%%%%%%%%%%%%%%%%%%%
%%  Datei: laztwist.tex  %%
%%%%%%%%%%%%%%%%%%%%%%%%%%%

\section{Twist decomposition of nonlocal LC operators}
\renewcommand{\theequation}{\thesection.\arabic{equation}}
\setcounter{equation}{0}
\label{twist}

This Chapter is devoted to show how the LC operators may be
decomposed according to their twist being defined by the rule
twist$(\tau)$ = dimension$(d)$ -- spin$(j)$. The procedure 
to be used is the following:
\begin{itemize}
\item
first, for {\em arbitrary} values of $x$
we {\em expand} the nonlocal operators  into a 
{\em Taylor series of local tensor operators} having 
definite rank and mass dimension,
\item
second, we {\em decompose} these tensor operators with respect to
{\em irreducible representations} of the Lorentz group or,
equivalently, the group $SO(4)$  having definite spin
and therefore also well-defined twist,
\item
third, we {\em resum} the infinite series of irreducible 
tensor operators 
of equal twist (for any $n$) to a {\em nonlocal 
harmonic operator of definite twist}
which contains, through its trace terms, also infinitely many 
operators of higher twist related to it, and
\item
finally, we {\em project onto the light--cone},
$x\rightarrow\tilde{x}$, to obtain the required
twist decomposition of the original light--cone operators.
\end{itemize}
The nonlocal quark operators to be considered
and their Taylor expansions are given by:
\begin{align}
\label{O_Gamma}
\hspace{-.2cm}
O^\Gamma(\kappa_1 \tilde x,\kappa_2\tilde{x})
&=
\bar{\psi}(\kappa_1 \tilde x)\Gamma 
U(\kappa_1 \tilde x,\kappa_2\tilde{x})
\psi(\kappa_2\tilde{x})
\\
\hspace{-.2cm}
\label{O_ent}
&=
\lim_{x\to\tilde{x}}
\sum_{n=0}^{\infty}
\frac{1}{n!}
{x}^{\mu_1}\ldots {x}^{\mu_n}
\bar{\psi}(0)\Gamma
{\sf D}_{\mu_1}\!(\kappa_1, \kappa_2)
\ldots 
{\sf D}_{\mu_n}\!(\kappa_1, \kappa_2)
\psi (0) ,
\end{align}
\vspace{-.2cm}
where the phase factor 
$U(\kappa_1\tilde{x},\kappa_2\tilde{x})$
has been defined by eq.~(\ref{phase}) and 
\begin{eqnarray}
\Gamma = \{1,  \gamma_\mu, \sigma_{\mu\nu}, 
%\; {\rm resp.}\; \ii\gamma_5\sigma_{\mu\nu} ,
\gamma_5 \gamma_\mu, \gamma_5\}
\quad {\rm with} \quad
\sigma_{\mu\nu} = 
\hbox{\large$\frac{i}{2}$}
[\gamma_\mu, \gamma_\nu ]
=
\hbox{\large$\frac{i}{2}$}
{\epsilon_{\mu\nu}}^{\kappa\lambda}
\gamma_5 \sigma_{\kappa\lambda}
\nonumber
\end{eqnarray}
determines a specific $\gamma$--structure (and, if necessary, 
also the flavour content) of the operators
under consideration\footnote{
The symmetry with respect to the exchange of $\kappa_1,
\kappa_2$ 
and the renormalization procedure of the LC operators 
have not been explicitly indicated since they do not play any 
role in our consideration here; for corresponding details see
\cite{BGR98}.
}; 
furthermore we have introduced the notation
\begin{align}
\label{D_kappa}
{\sf D}_\mu(\kappa_1, \kappa_2)
&\equiv 
(\kappa_1\!\! \stackrel{\leftarrow}{D}
+
\kappa_2\!\! \stackrel{\rightarrow}{D})_{\mu},\\
{\rm with} \qquad\qquad
{\stackrel{\rightarrow}{D}}_{\mu}
\equiv
D_\mu(A)=\pd_\mu &+\ii g A_\mu(x)
\quad{\rm and}\quad
{\stackrel{\leftarrow}{D}}_\mu
= {\stackrel{\leftarrow}{\pd}}_\mu-\ii g A_\mu(x).
\end{align}
Obviously, the expansion (\ref{O_ent}) of 
$O^\Gamma(\kappa_1, \kappa_2)$ 
into a Taylor series makes use of the translation %property 
of the field operators, 
$\psi(\kappa x)=\exp(\ii\kappa xP)\psi(0)\exp(-\ii\kappa xP)$,
together with
$[P_{\mu},\psi(0)] = - \ii(\pd_{\mu}\psi)(0)$.
%\pagebreak

%An analogous expansion obtains for the gluon operators:
%\begin{align}
%\label{OV_ent}
%\hspace{-.2cm}
%O^V_{\mu\nu}(\kappa_1, \kappa_2)
%&=
%F^a_{\mu\lambda}(\kappa_1 \tilde x)
%U^{ab}(\kappa_1 \tilde x,\kappa_2\tilde{x})
%F^{b\lambda}_\nu(\kappa_2\tilde{x})
%\\
%\hspace{-.2cm}
%&=
%\lim_{x\to\tilde{x}}
%\sum_{n=0}^{\infty}
%\frac{1}{n!}
%{x}^{\mu_1}\ldots {x}^{\mu_n}
%F^a_{\mu\lambda}(0)
%{\sf D}^{ac_1}_{\mu_1}(\kappa_1, \kappa_2)
%\ldots 
%{\sf D}^{c_{n-1}b}_{\mu_n}(\kappa_1, \kappa_2)
%F^{b\lambda}_\nu(0),
%\nonumber
%\\
%\label{OA_ent}
%\hspace{-.2cm}
%O^A_{\mu\nu}(\kappa_1, \kappa_2)
%&=
%F^a_{\mu\lambda}((\kappa_1 \tilde x)
%U^{ab}(\kappa_1 \tilde x,\kappa_2\tilde{x})
%{\tilde F}^{b\lambda}_\nu(\kappa_2\tilde{x})
%%\nonumber
%\\
%\hspace{-.2cm}
%&=
%\lim_{x\to\tilde{x}}
%\sum_{n=0}^{\infty}
%\frac{1}{n!}
%{x}^{\mu_1}\ldots {x}^{\mu_n}
%F^a_{\mu\lambda}(0)
%{\sf D}^{ac_1}_{\mu_1}(\kappa_1, \kappa_2)
%\ldots 
%{\sf D}^{c_{n-1}b}_{\mu_n}(\kappa_1, \kappa_2)
%{\tilde F}^{b\lambda}_\nu(0),
%\nonumber
%\end{align}
%where $F_{\mu\nu} = F^a_{\mu\nu}t^a$ and 
%${\tilde F}_{\mu\nu} = \frac{1}{2} 
%{\epsilon_{\mu\nu}}^{\kappa\lambda} F_{\kappa\lambda}$
%is the gluon field strenght and its dual, respectively,
%which, together with the phase factor, are taken in the 
%adjoint representation.

The local operators appearing in eqs.~(\ref{O_ent}) 
%(\ref{OV_ent}) and (\ref{OA_ent}) 
have a tensor structure which, in general, is reducible with
respect to the (orthochronous) Lorentz group. However, since
only the tensor structure of these operators is of interest,
we may restrict the following considerations to the much
simpler case $(\kappa_1 = 0, \kappa_2 = \kappa)$ where
${\sf D}_{\mu}(\kappa_1,\kappa_2)$
reduces to the usual covariant derivative $D_\mu$ multiplied 
by $\kappa$. Any (algebraic) operation which will be used henceforth
is unchanged by the replacement
$\kappa D_\mu \rightarrow {\sf D}_{\mu}(\kappa_1,\kappa_2)$.
This is independent from the fact that the product
${\sf D}_{\mu_1}\!(\kappa_1, \kappa_2)\ldots 
{\sf D}_{\mu_n}\!(\kappa_1, \kappa_2)$, if expanded according
to (\ref{D_kappa}), decays into a sum of tensors -- although
different -- being equivalent realizations of the same
symmetry types.
Therefore, the twist decomposition which will be obtained
for 
$O^\Gamma(0, \kappa)$ immediately holds with obvious
replacements for
$O^\Gamma(\kappa_1, \kappa_2)$.
%, and similarly for the gluon operators.

There is an additional property of the considered operators.
Since the product of the covariant derivatives is multiplied
by the symmetric tensor ${x}^{\mu_1}\ldots {x}^{\mu_n}$ only
its symmetric part $D_{\{\mu_1}\ldots D_{\mu_n\}}$ 
is of relevance where the symbol $\{\ldots\}$
denotes symmetrization with respect to the enclosed indices
%appearing within the bracket 
(including division by $n!$); in the same manner
$[\ldots]$ will be used to denote antisymmetrization.
Of course, $D_{\{\mu_1}\ldots D_{\mu_n\}}$ as any 
 symmetric tensor decomposes 
into irreducible tensors according to the Clebsch--Gordan series
\begin{eqnarray}
\label{totsym}
D_{\{\mu_1}\ldots D_{\mu_n\}}\in
{\bf T}\Big(\hbox{\large$\frac{n}{2},\frac{n}{2}$}\Big)
\oplus
{\bf T}\Big(\hbox{\large$\frac{n-2}{2},\frac{n-2}{2}$}\Big)
\oplus
{\bf T}\Big(\hbox{\large$\frac{n-4}{2},\frac{n-4}{2}$}\Big)
\oplus\ldots,
%\nonumber
\end{eqnarray}
where the traceless tensor of order $n$ transform according to
${\bf T}(\hbox{$\frac{n}{2},\frac{n}{2}$})$, and any of the 
further representations correspond to the traces which
contain an increasing number of metric tensors (multiplied
by $x^2$).
Therefore, in the limit $x \rightarrow \tilde x$ only the
tensor space ${\bf T}(\hbox{$\frac{n}{2},\frac{n}{2}$})$
contributes. This leads to essential restrictions for the
decomposition of the LC--operators under consideration. 

The indices resulting from the Taylor decomposition of the
LC--operators will be called ``internal'' ones, whereas the 
indices which are due to the nonlocal operator itself,
i.e.,~which are related to $\Gamma$ will be
called ``external''.
For some physical situations it will be useful to multiply
also (some of the) external indices by 
$x^\mu$ and ${\tilde x}^\mu$, respectively,
thereby extending the above conclusion.
%
%%%%%%%%%%%%%%%%%%%%%%%%%%%%%%%%%%%%%%%
%%%%%
\subsection{(Pseudo) Scalar operators}
\label{scalar}
%%%%%%%%%%%%%%%%%%%%%%%%%%%%%%%%%%%%%%%
%%%%%
%
The simplest cases to demonstrate how the method works are the 
(pseudo) scalar operators. In order to think of some specific example we 
could take $\Gamma = \{1, (\xx\gamma)=\xx^\mu \gamma_\mu, 
(\xx\sigma\partial)=\xx^\mu\sigma_{\mu\nu}\partial^\nu,  
\gamma_5(\xx\gamma), \gamma_5\}$ leading to operators having
lowest twist  $\tau_{\rm min}= \{3, 2, 2, 2, 3\}$, respectively.
Here, any Lorentz index of the local operators is an internal one and
only the Young tableaux of type (i) -- 
corresponding to complete symmetrisation -- 
are of relevance.  According to eq.~(\ref{totsym}) 
they contain contributions of
twist $\tau_n = \tau_{\rm min} + 2n, n = 0,1,2,\ldots $ 
which, on the light--cone,
reduce to $\tau_{\rm min}$ only.

To be definite, let us consider the operator 
\bea
\label{O2}
O(0,\kappa x)
=
\bar{\psi}(0) \lcxg 
U(0,\kappa x)\psi(\kappa x)\,,
\eea
which also appears in the consideration of vector operators.
Its decomposition into local tensor operators reads
\bea
\label{O2loc}
%O(0,\kappa x)=
\sum_{n=0}^\infty \frac{\kappa^n}{n!}
x^{\beta}x^{\mu_1}\ldots x^{\mu_n}
\bar{\psi}(0) 
\gamma_\beta D_{\mu_1}\ldots D_{\mu_n}
\psi(0)
=
\sum_{n=0}^\infty \frac{\kappa^n}{n!}
\bar{\psi}(0) \lcxg (xD)^n
\psi(0)\, ,
\eea
with the obvious abbreviation
$(x D)^n
\equiv 
x^{\mu_1}\ldots x^{\mu_n} D_{\mu_1}\ldots D_{\mu_n}$.

Now, the local operators 
$\bar{\psi}(0) \gamma_\beta D_{\mu_1}\ldots D_{\mu_n}\psi(0)$ 
have to be made traceless. This is easily achieved if we 
observe that a totally symmetric traceless tensor 
$\tl T_{\mu_1\ldots\mu_n}$ whose indices are completely 
saturated by contracting with $x^{\mu_1}\ldots x^{\mu_{n}}$
is a harmonic polynomial of degree $n$. 
It obeys the 4--dimensional
Laplace equation if continued to Euclidean space--time,
\begin{equation}
\label{Bed_tw2}
\square \tl T_n(x)=0
\quad {\rm with}\quad
\tl T_n(x) := x^{\mu_1}\ldots x^{\mu_{n}}
\tl T_{\mu_1\ldots\mu_n},
\end{equation}
which is obvious if one observes that
$\square x^\mu x^\nu \equiv 2 g^{\mu\nu}$.
%\\
The solution of eq.~(\ref{Bed_tw2}) is given by
(see Appendix~\ref{trace}, eq.~(\ref{T_harm4})):
\begin{equation}
\label{Proj_tw2}
\tl T_n (x)
=
\sum_{k=0}^{[\frac{n}{2}]}\frac{(n-k)!}{n!k!}
\left(\frac{-x^2}{4}\right)^{\!k}\square^k
 x^{\mu_1}\ldots x^{\mu_n} T_{\mu_1\ldots\mu_n}
\equiv
H_n^{(4)}\!\left(x^2|\square\right) T_n(x)\ .
\end{equation}
From this we conclude, after replacing
$n \rightarrow n+1$, that the local traceless quark operators 
in eq.~(\ref{O2loc}) read
\begin{equation}
\left[ {\bar\psi}(0)\lcxg (xD)^n \psi(0)\right]^{\rm traceless}
=
\sum_{k=0}^{[\frac{n+1}{2}]}
\frac{(n+1-k)!}{(n+1)!k!}
\left(\frac{-x^2}{4}\right)^{\!k} \square^k\,
\bar{\psi}(0) \lcxg (xD)^n \psi(0).
\nonumber
\end{equation}
Let us  remark that the summation over $k$ could be left unbounded
since, due to the $k$--th power of the Laplacian, it automatically
terminates at $[\frac{n+1}{2}]$.

Now, these traceless operators of  twist--2  must be resummed
by introducing  them into eq.~(\ref{O2loc}).  Using 
 the integral representation of 
Euler's beta function,
\begin{equation}
\label{Euler}
B(n,m)
=\frac{\Gamma(n)\Gamma(m)}{\Gamma(n+m)}
=\int_0^1\d t\, t^{n-1}(1-t)^{m-1},
\end{equation}
in order to replace ${(n+1-k)!}/{(n+1)!}
= B(n+2-k,k) /(k-1)!$ we obtain the following
traceless operator
\bea
%\lefteqn{\tl O(0,\kappa x)}
%\nonumber\\
%&=&
\sum_{n=0}^\infty
\bigg[1+\sum_{k=1}^\infty\int_0^1\d t
\left(\frac{-x^2}{4}\right)^{\!k}
\frac{\square^k\,t^n}{k!(k-1)!}
\left(\frac{1-t}{t}\right)^{\!k-1}\bigg]
\frac{\kappa^n}{n!}\bar{\psi}(0) \lcxg (x D)^n \psi(0).
\nonumber
\eea
Here, the sum over $n$ can be carried out leading to 
\begin{align}
\label{proj_tw2}
%\lefteqn{
\tl O(0,\kappa x)
=&\;
\bar{\psi}(0)\lcxg U(0,\kappa x) \psi (\kappa x)%}
%\nonumber
\\
& 
+\sum_{k=1}^{\infty}\int_0^1\!\!\d t
\left(\frac{-x^2}{4}\right)^{\!k}\!\!
\frac{\square^k}{k!(k-1)!}
\left(\frac{1-t}{t}\right)^{\!\! k-1}
\bar{\psi}(0)\lcxg U(0,\kappa tx) \psi (\kappa tx)\ .
\nonumber
\end{align}
This expression already has been dealt with by \cite{BB88}
without mentioning its derivation.
It is obvious that the terms corresponding to $k = 1, 2,\ldots$
contain operators of higher twist $\tau = 4, 6, \ldots$ which,
because they are multiplied with $x^2$, do not contribute on
the light--cone. Therefore, the scalar light--cone operator
$\bar{\psi}(0) \gamma\xx U(0,\kappa\xx) \psi(\kappa\xx)$
is already of twist--2.

Let us observe that the derivation of (\ref{proj_tw2}) is independent 
of the origin of the specific structure of the local tensors appearing
in eq.~(\ref{O2loc}). Therefore, we may replace $\kappa D_\mu$ by
${\sf D}_\mu(\kappa_1, \kappa_2)$ without changing the final
result:
\bea
\label{proj_tw2k}
\lefteqn{\tl O(\kappa_1 x, \kappa_2 x)
=
\bar{\psi}(\kappa_1 x)\lcxg U(\kappa_1 x,\kappa_2 x) \psi (\kappa_2 x)
}
%\nonumber
\\
& &
+\sum_{k=1}^{\infty}\int_0^1\d t
\left(\frac{-x^2}{4}\right)^{\!k}
\frac{\square^k}{k!(k-1)!}
\left(\frac{1-t}{t}\right)^{k-1}
\bar{\psi}(\kappa_1 tx)\lcxg U(\kappa_1 tx,\kappa_2 tx) \psi (\kappa_2 tx)\ .
\nonumber
\eea
The projection onto the light--cone leads to
\bea
\label{O2k}
O^{\rm tw2}(\kappa_1 \xx, \kappa_2 \xx)
&=&
\bar{\psi}(\kappa_1 \xx) (\xx\gamma) 
U(\kappa_1 \xx,\kappa_2 \xx) \psi (\kappa_2 \xx);
\eea
of course, any of the scalar structures $\Gamma$
introduced above could be used instead of $(\xx\gamma)$.
%
%
%%%%%%%%%%%%%%%%%%%%%%%%%%%%%%%%%%%%%%%
%%%%%
\subsection{(Pseudo) Vector operators}
\label{vector}
%%%%%%%%%%%%%%%%%%%%%%%%%%%%%%%%%%%%%%%
%%%%%
%
Let us now consider the (pseudo) vector
operators with $\Gamma$ being $\gamma_\beta$ (or 
$\gamma_\beta\gamma_5$):
\bea
\label{Ovector}
O_\beta(0,\kappa x)
=
{\bar\psi}(0)\gamma_\beta 
U(0,\kappa x)\psi(\kappa x)\, ,
\eea
whose local tensor operators  differ
from those in eq.~(\ref{O2loc}) only by
the absence of $x^\beta$. They are given by
\begin{eqnarray}
\hspace{-1.2cm}
\label{O_loc}
O_{\beta\mu_1\ldots\mu_n}
\!\!\!&\equiv&\!\!\!
\bar{\psi}(0)\gamma_\beta D_{\{\mu_1}\ldots D_{\mu_n\}}\psi(0)
\\
\hspace{-1.2cm}
\label{O_loc1}
\!\!\!&{=}&\!\!\!
\bar{\psi}(0)
\gamma_{\{\beta} D_{\mu_1}\ldots D_{\mu_n\}}
\psi(0)
+
\alpha_n
%\hbox{\large$\frac{2n}{n+1}$}
\bar{\psi}(0)
\gamma_{[\beta} D_{\{\mu_1]}\ldots D_{\mu_n\}}
\psi (0)
%\\
%\hspace{-1cm}
%\!\!\!&&\!\!\!
+ \ldots ;
%\nonumber
\end{eqnarray}
they decompose into two parts being related to the
Young patterns (i) and (ii) as well as to
additional contributions being
antisymmetric in (some of) the $\mu$'s, and
 $\alpha_n = 2n/(n+1)$ results from the 
nontrivial normalization of the Young operator 
${\cal Y}_{[m]}$ with $[m] = (n,1)$
(see Appendix \ref{young}, eqs.~(\ref{Young1})
and (\ref{f})).
%(The undefined coefficient $\alpha_n$ %in eq.~(\ref{O_loc1}) 
%is determined by eq.~(\ref{prefac}) below).
Let us remind that because of the
full reducibility of the direct product of a vector with
a completely symmetric tensor (\ref{O_loc}) corresponds
 to the following Clebsch-Gordan series
\begin{eqnarray}
\label{CGR1}
\lefteqn{
\hbox{\large$\big(\frac{1}{2},\frac{1}{2}\big)$}
\otimes\left(
\hbox{\large$\big(\frac{n}{2},\frac{n}{2}\big)$}
\oplus
\hbox{\large$\big(\frac{n-2}{2},\frac{n-2}{2}\big)$}
\oplus
\ldots\right)}
\\
&&
=\hbox{\large$\big(\frac{n+1}{2},\frac{n+1}{2}\big)$}
\oplus
\left(\hbox{\large$\big(\frac{n+1}{2},\frac{n-1}{2}\big)$}
\oplus
\hbox{\large$\big(\frac{n-1}{2},\frac{n+1}{2}\big)$}\right)
\oplus
2 \hbox{\large$\big(\frac{n-1}{2},\frac{n-1}{2}\big)$}
\oplus\ldots\ .\qquad 
\nonumber
\end{eqnarray}
The mass dimension of the operator (\ref{Ovector}) is $n+3$, 
and the spin of the various contributions in (\ref{CGR1}) 
ranges from $n+1$ up to $1$ or $0$ if $n$ is even or odd,
respectively; therefore the twist decomposition reads:
\begin{equation} 
\label{O_tw}
 O_{\beta\mu_1\ldots\mu_n}
=O^{\mathrm{tw2}}_{\beta\mu_1\ldots\mu_n}
+O^{\mathrm{tw3}}_{\beta\mu_1\ldots\mu_n}
+O^{\mathrm{tw4}}_{\beta\mu_1\ldots\mu_n}
+\ldots \;,
\end{equation}
with the last term in the series being
$O^{\mathrm{tw(n+2)}}_{\beta\mu_1\ldots\mu_n}$ or
$O^{\mathrm{tw(n+3)}}_{\beta\mu_1\ldots\mu_n}$, respectively.
From this it follows that the operator (\ref{O_ent}) for arbitrary
$x$ contains contributions of any twist, $\tau \geq 2$. 
However, in the limit
$x \rightarrow {\tilde x}$ only the first three terms 
of eq.~(\ref{O_tw}) survive
since the contributions for $\tau > 4$ are proportional 
to $x^2$. Therefore we obtain the following decomposition:
\begin{equation} 
\label{O_tw_nl}
O_{\beta}(0,\kappa\tilde{x})
=
 O^{\mathrm{tw2}}_{\beta}(0,\kappa\tilde{x})
+O^{\mathrm{tw3}}_{\beta}(0,\kappa\tilde{x}) 
+O^{\mathrm{tw4}}_{\beta}(0,\kappa\tilde{x}) .
\end{equation}
On the light--cone the totally symmetric part of eq.~(\ref{O_loc1}) 
contributes to $\tau = 2, 4 $,
and the partially antisymmetric part of (\ref{O_loc1}) 
contributes to $\tau = 3, 4$. 
%Since the two series of irreducible
%representations are independent from each other 
Obviously, the twist--4 operator consists of two independent parts.
%Let us now consider the two different symmetry classes
%(i) and (ii) separately.

\smallskip
\noindent
(A)~~~{\em Totally symmetric vector operators}:\\
 The totally symmetric traceless 
tensors which have twist $\tau =2$ are contained in 
${\bf T} (\frac{n+1}{2},\frac{n+1}{2})$ and they are 
characterized by the following standard tableaux:
\\
\\
%\vspace*{3mm}
\unitlength0.5cm
\begin{picture}(30,1)
\linethickness{0.075mm}
\put(1,0){\framebox(1,1){$\beta$}}
\put(2,0){\framebox(1,1){$\mu_1$}}
\put(3,0){\framebox(1,1){$\mu_2$}}
\put(4,0){\framebox(3,1){$\ldots$}}
\put(7,0){\framebox(1,1){$\mu_n$}}
\put(10,0){$\stackrel{\wedge}{=}$}
\put(12.5,0){$\relstack{\beta\mu_1\ldots\mu_n}{\cal S}
\bar{\psi}(0)\gamma_{\beta} D_{\mu_1}\ldots D_{\mu_n}\psi(0)
- \mathrm{trace~terms}$}
\end{picture}
\\
\\
where the normalizing factor is 1. Let us write this operator in the form
\begin{align}
\label{O2_loc}
O^{\mathrm{tw2}}_{\beta\mu_1\ldots\mu_n}
=&
\hbox{\large$\frac{1}{n+1}$}
\Big(
\bar{\psi}(0)\gamma_{\beta} D_{\{\mu_1}\ldots
D_{\mu_n\} } \psi(0)
\\ 
&+\!
\sum_{l=1}^{n} \bar{\psi}(0)\gamma_{\{\mu_l}
D_{\mu_1}\ldots D_{\mu_{l-1}}{D}_{|\beta|}
D_{\mu_{l+1}}\ldots D_{\mu_n\} }\psi(0)\!\Big)\!
-\!\mathrm{trace~terms} ,%\qquad
\nonumber 
\end{align}
where the symbol $|\ldots|$ indicates that the enclosed
indices are to be excluded from the symmetrization caused
by $\{\ldots\}$. If the trace terms in (\ref{O2_loc}) 
had been determined, 
this tensor would be irreducible. Let us postpone
that determination and make the resummation to the 
corresponding nonlocal operator in advance. This is obtained
by contracting with $x^{\mu_1}\ldots x^{\mu_n}$:
\begin{eqnarray}
\label{2betan}
\hspace{-.2cm}
O^{\mathrm{tw2}}_{\beta n}(x)
=%&:=&
x^{\mu_1}\ldots
x^{\mu_n}O^{\mathrm{tw2}}_{\beta\mu_1\ldots\mu_n}
%\nonumber\\
\label{O2_x}
=%&=&
\hbox{\large$\frac{1}{n+1}$}
\pd_\beta\Big(
\bar{\psi}(0)\lcxg (xD)^n \psi(0)
-\mathrm{trace~terms}\Big).
\end{eqnarray}
The trace terms of $\bar{\psi}(0)\lcxg (xD)^n \psi(0)$ have been
determined already in Subsection \ref{scalar}. Therefore, 
using $(n+1)^{-1} = \int^1_0 d\lambda \lambda^n$ we obtain:
\bea
\label{O_tw2}
%\label{O2_beta}
O^{\mathrm{tw2}}_{\beta} (0,\kappa x)
=
\sum_{n=0}^{\infty}
\frac{\kappa^n}{n!} 
O^{\mathrm{tw2}}_{\beta n}(x)
=
\pd_\beta
\int_{0}^{1} \d\lambda\,
\tl O(0,\kappa\lambda x) ,
\vspace{-.3cm}
\eea
with $\tl O(0,\kappa\lambda x)$ being defined through
eq.~(\ref{proj_tw2}).
Of course, the derivative $\pd_\beta$ could be taken outside the 
parentheses in eq. (\ref{2betan}) (and the integral) 
because the Young operators should be applied
to traceless tensors. (For notational definiteness we remark
that partial derivatives are everywhere with respect to
$x$ only.)
Note that $x^\beta O^{\mathrm{tw2}}_{\beta} (0,\kappa x)=
\tl O(0,\kappa x)$ is obtained 
from eq.~(\ref{O_tw2}) by partial integration 
observing that for any function $f(\lambda x)$ 
the equality $x^\mu\partial f/\partial x^\mu =
\lambda \partial f/\partial \lambda$ holds. 
Analogous to the scalar case the
generalization to arbitrary values $(\ka, \kb)$ is 
%simply obtained as
\bea
\label{O2_betak}
O^{\mathrm{tw2}}_{\beta} (\ka x,\kb x)
=
\pd_\beta
\int_{0}^{1} \d\lambda\,
\tl O(\ka\lambda x,\kb\lambda x) ,
\vspace{-.3cm}
\eea
%\vspace{-.3cm}
with $\tl O(\ka\lambda x,\kb\lambda x)$ defined through eq.
(\ref{proj_tw2k}).
By construction this traceless harmonic vector operator 
fulfills the equations
\bea
\label{cond}
\square O^{\mathrm{tw2}}_{\beta}  (\ka x,\kb x)=0\ ,
\qquad
\pd^\beta O^{\mathrm{tw2}}_{\beta}  (\ka x,\kb x)=0\, .
\eea

Let us now project onto the light--cone.
Since only one differentiation $\pd_\beta$
appears in eq.~(\ref{O_tw2}) and (\ref{O2_betak})
 the whole sum in (\ref{proj_tw2}) and (\ref{proj_tw2k}), 
respectively, contributes on the light--cone only
through the term $k=1$.
Using the relation
\begin{equation}
\int_{0}^{1}\d\lambda\int_{0}^{1} \d t f(\kappa\lambda t)
=-\int_0^1\d \lambda(\ln\lambda)f(\kappa \lambda)
\end{equation}
we finally obtain
for the nonlocal twist--2 light--cone operator
\begin{align}
\hspace{-.2cm}
\label{Otw2_gir}
O^{\mathrm{tw2}}_{\beta}(\ka\xx,\kb\xx)
&=
\lim_{x\to\tilde{x}}
\pd_\beta\!\!
\int_{0}^{1} \d\lambda
%\left\{
\left[1+\hbox{\large$\frac{\ln\lambda}{4}$}x^2\square\right]
\bar{\psi}(\ka\lambda x)\lcxg 
U(\ka\lambda x,\kb\lambda x) 
\psi(\kb\lambda x)%\right\}
\nonumber \\
&=\!
\int_{0}^{1}\!\! \d\lambda\!\!
\left[\pd_\beta +\hbox{\large$\frac{\ln\lambda}{2}$}
x_\beta\square\right]\!\!
\left.\bar{\psi}(\ka\lambda x)\lcxg 
U(\ka\lambda x,\kb\lambda x) 
\psi(\kb\lambda x)\right|_{x=\tilde{x}}\!.\!\!
\end{align}
Obviously, the second term of the integrand which is 
proportional to $x_\beta$, and which results from $k=1$, is the 
trace term; up to an overall minus sign it is the (symmetric)
twist--4 vector operator
\bea
\label{O4s}
\hspace{-.4cm}
O^{\mathrm{tw4}}_{\beta, {\rm sym}} (\ka\xx,\kb\xx)
=
- \hbox{\large$\frac{1}{2}$}\xx_\beta\square\!\!
\int_{0}^{1}\!\! \d\lambda\,\ln\lambda
\left.\bar{\psi}(\ka\lambda x)\lcxg 
U(\ka\lambda x,\kb\lambda x) 
\psi(\kb\lambda x)\right|_{x=\tilde{x}}\!.
\eea
Of course, if we multiply (\ref{Otw2_gir}) with $\xx^\beta$
the twist--4 part vanishes and the remaining contributions 
restore the scalar operator (\ref{O2k}).

Furthermore, after performing the differentiations in 
(\ref{Otw2_gir}) we can use the axial gauge, 
$\xx^\mu A_\mu = 0$, in order to avoid the phase factor,
$U(\ka\xx,\kb\xx)= 1$; then we obtain
%\begin{eqnarray}
%\hspace{-.5cm}
%\label{Otw2_irax}
%O^{\mathrm{tw2}}_{\beta}(\ka\xx,\kb\xx)
%=
%\!\int_{0}^{1} \!\d\lambda \,\bar{\psi}(0)\Big(
%\gamma_\beta +\lcyg D_\beta%(y) %\right]
%+ y_\beta \ln\lambda
%\left.
%\!\!\left[  D\Dslas+ \hbox{\large$\frac{1}{2}$}
%\lcyg  D^2\right]\!\!\Big)
%\psi(y)\right|_{y=\kappa\lambda\xx},
%\end{eqnarray}
%where the covariant derivatives are taken with respect to
%the argument $y$. Its generalization 
%to arbitrary values of $(\ka, \kb)$ reads
\begin{eqnarray}
\hspace{-.5cm}
\label{Otw2_iraxk}
O^{\mathrm{tw2}}_{\beta}(\ka\xx,\kb\xx)
=
\!\int_{0}^{1} \!\d\lambda \,\bar{\psi}(\ka y)\Big(
\gamma_\beta +\lcyg {\sf D}_\beta
+ y_\beta \ln\lambda
\left.
\!\!\left[ {\sf D}\slass + \hbox{\large$\frac{1}{2}$}
\lcyg {\sf D}^2\right]\!\!\Big)
\psi(\kb y)\right|_{y=\lambda\xx},
%\nonumber
\end{eqnarray}
with the {\em convention} that the covariant 
left--right derivative in axial gauge,
\begin{eqnarray}
{\sf D}_\beta
\equiv 
\overleftarrow{\partial_\beta^y}
-ig\int_{\kb}^{\ka} d\tau A_\beta(\tau y) 
+\overrightarrow{\partial_\beta^y},
\nonumber
\end{eqnarray} 
acts only on the arguments of the quark and gluon fields. (This
convention which is used also in the following, is 
necessary only to simplify the notation.)

Finally, it should be remarked, that the conditions
(\ref{cond}), if translated into the corresponding ones
with respect to $\xx$, no longer hold for the light--cone operators
(\ref{Otw2_gir}). This is obvious because by projecting onto the
light--cone part of the original structure of the operators has been lost.
Nevertheless, the resulting light--cone operators are restrictions of
harmonic functions. This was the reason why in order to determine
the correct operators of definite twist we started with the 
quark operators for arbitrary values of $x$.

\smallskip
%%%%%%%%%%%%%%%%%%%%%%%%%%%%%%%%%%%%%%%
\noindent
(B)~~~{\em Vector operators having symmetry class (ii)}\\
Now we consider the symmetry class (ii) which contributes to 
 spin $j=n$ and $n-1$. 
Because of the specific symmetry of the local
operators (\ref{O_loc}) there appears only one nontrivial 
standard tableaux:\footnote{
Local operators with this symmetry have been considered first 
by~\cite{Ahm76,Kod79} in connection with the polarized
deep inelastic scattering.}
\\
\\
\unitlength0.5cm
\begin{picture}(30,1)
\linethickness{0.075mm}
\put(1,-1){\framebox(1,1){$\beta$}}
\put(1,0){\framebox(1,1){$\mu_1$}}
\put(2,0){\framebox(1,1){$\mu_2$}}
\put(3,0){\framebox(1,1){$\mu_3$}}
\put(4,0){\framebox(3,1){$\ldots$}}
\put(7,0){\framebox(1,1){$\mu_n$}}
\put(8.5,0){$\stackrel{\wedge}{=}$}
\put(9.5,0)
{${\hbox{\large$\frac{2n}{n+1}$}}\,
\relstack{\beta\mu_1}{\cal A}\,
\relstack{\mu_1\ldots\mu_n}{\cal S}\!\!
\bar{\psi}(0) \gamma_{\beta} 
D_{\mu_1}\ldots D_{\mu_n}\psi(0)
-\mathrm{trace~terms}$}
\end{picture}
\\ 
\\
with the normalization $\alpha_n = 2n/(n+1)$ of the 
Young operator.
The corresponding traceless tensor being contained
in ${\bf T}(\frac{n+1}{2},\frac{n-1}{2})\oplus
{\bf T}(\frac{n-1}{2},\frac{n+1}{2})$
has twist $\tau = 3$; it is reducible with respect to
$SO(1,3)$, but irreducible with respect to $O(1,3)$:
\begin{eqnarray}
\label{O3_loc}
O^{\mathrm{tw3}}_{\beta\mu_1\ldots\mu_n}
=
\hbox{\large$\frac{n}{n+1}$}
\bar{\psi}(0)\Big(\gamma_{\beta} D_{\{\mu_1}\ldots
D_{\mu_n\} } %\psi(0)
- %\bar{\psi}(0)
\gamma_{\mu_1} D_{\{\beta}\ldots D_{\mu_n\} }
\Big)\psi(0)
%\nonumber\\
%& &
-\mathrm{trace~terms}.
\nonumber
\end{eqnarray}
This irreducible tensor will be contracted with 
$x^{\mu_1}\ldots x^{\mu_n}$ to obtain:
\begin{eqnarray}
O^{\mathrm{tw3}}_{\beta n}(x)
%&=&x^{\mu_1}\ldots
%x^{\mu_n}O^{\mathrm{tw3}}_{\beta\mu_1\ldots\mu_n}\\
\label{O3x_loc}
=
\hbox{\large$\frac{2}{n+1}$}
\hbox{\large$\frac{1}{2}$}
\Big(g_{\alpha\beta}(x\pd)-x_\alpha\pd_\beta\Big)
\bar{\psi}(0)\gamma^\alpha (x D)^n \psi (0)
-\mathrm{trace~terms}\ .
%\nonumber
\end{eqnarray}
%Here, some remarks are in order.
Obviously, the differential operator
$\frac{1}{2}
\left(g_{\beta\alpha}(x\pd)-x_\alpha\pd_\beta\right)$
takes care of the antisymmetrization
${\cal A}_{\beta\mu_i}$ after having symmetrized by 
contracting with the $x$'s. However, this contraction
in addition causes that those contributions to
(\ref{O_loc1}) which are antisymmetric in some of the
$\mu$'s vanish such that the following relation holds
\begin{equation}
\hspace{-.2cm}
\label{prefac}
x^{\mu_1}\ldots x^{\mu_n}O_{\beta\mu_1\ldots\mu_n}
=
x^{\mu_1}\ldots x^{\mu_n}\Big(
\relstack{\beta\mu_1\ldots\mu_n}{\cal S}\!\!\!
O_{\beta\mu_1\ldots\mu_n}
+
\hbox{\large$\frac{2n}{n+1}$}
\relstack{\beta\mu_1}{\cal A}\,
\relstack{\mu_1\ldots\mu_n}{\cal S}\!\!\!
O_{\beta\mu_1\ldots\mu_n}\Big).
%\nonumber
\end{equation}
Of course, this relation could have been used to determine
the prefactor $\alpha_n$ without any knowledge of the
normalization of the Young operators.\footnote{
\label{SYM}
In fact, contraction with the $x$'s could be understood
as a supplementary symmetrization 
according to ${\cal S}_{\mu_1\ldots\mu_n}
{\cal A}_{\beta\mu_1}{\cal S}_{\mu_1\ldots\mu_n}$ which
has been first considered by~\cite{Shu82,Bukh83}.
It corresponds to taking into account all the $n!$ 
equivalent representations related to the above Young pattern 
or simply replacing
$\gamma_{\mu_1} D_{\{\beta}\ldots D_{\mu_n\}}$
by 
$\frac{1}{n} \sum_\ell 
\gamma_{\mu_\ell} D_{\{\mu_1}\ldots D_{\mu_{\ell-1}}
D_\beta D_{\mu_{\ell+1}}\ldots D_{\mu_n\} }$.
Obviously,  this additional symmetry operation 
destroys the irreducibility and, furthermore, applied to
tensors being not traceless it may change the structure of the
trace terms. }

Again, before determining the trace terms we resum these
irreducible local tensors to the corresponding nonlocal operators:
\begin{align}
\hspace{-.3cm}
O^{\mathrm{tw3}}_{\beta} (0,\kappa x)
%=
%\sum_{n=0}^{\infty}\frac{\kappa^n}{n!} 
%O^{\mathrm{tw3}}_{\beta n}(x)%\nonumber \\
&=
\Big(\delta_{\beta}^{\alpha}(x\pd)-x^\alpha\pd_\beta\Big)
\!\!\int_{0}^{1} \!\!\d\lambda
\Big( \!O_\alpha(0, \kappa\lambda x)
-\mathrm{trace~terms}\!\Big).
\end{align}

In distinction to eq.~(\ref{Bed_tw2}) we now have to determine
a traceless tensor with a free index $\alpha$,
$\tl T_{\alpha n}(x)
\equiv
x^{\mu_1}\ldots x^{\mu_n} \tl T_{\alpha\mu_1\ldots\mu_n}$,
 obeying the following two equations:
\begin{equation}
\square \tl T_{\alpha n}(x)=0
\qquad{\rm and}\qquad
\pd^\alpha \tl T_{\alpha n}(x)=0,
\end{equation}
which correspond to taking the traces $g_{\mu_i\mu_j}$
and $g_{\alpha\mu_i}$, respectively.
The solution of these equations reads 
 (see Appendix~\ref{trace}, eq.~(\ref{Proj})):
\begin{equation}
\hspace{-.3cm}
\label{Proj1}
\tl T_{\alpha n}
=
\Big\{\delta_{\alpha}^{\mu}
-\hbox{\large$\frac{1}{(n+1)^2}$}
\big[x_\alpha\pd^\mu(x\pd)
%-x_\alpha x_\mu\square
- \hbox{\large$\frac{1}{2}$}
x^2\pd_\alpha\pd^\mu]\Big\}
\sum_{k=0}^{[\frac{n}{2}]}
\hbox{\Large$\frac{(n-k)!}{k!n!}$}
\left(\!\!\frac{-x^2}{4}\!\right)^{\!\!k}
\!\!\square^k T_{\mu{ n}}(x) .
\end{equation}
Again, using (\ref{Euler}) and the integral representation
of $1/(n+1)$ we obtain
\begin{align}
\label{proj_tw3}
\hspace{-.4cm}
\tl O_\alpha (&0,\kappa x)
=
O_\alpha(0,\kappa x)+\sum_{k=1}^\infty\int_0^1\!\!
\frac{\d t}{t}\!\!
\left(\!\frac{-x^2}{4}\!\right)^{\!k} \!\!
\frac{\square^k}{k!(k-1)!}
\left(\!\frac{1-t}{t}\!\right)^{\!\!k-1}\!\!
O_\alpha(0,\kappa t x)
%\nonumber
\\
%\hspace{-1cm}
&-
\big[ x_\alpha\pd^\mu(x\pd)
-\hbox{\large$\frac{1}{2}$}x^2\pd_\alpha\pd^\mu\big]
\sum_{k=0}^\infty\int_0^1\!\!\d\tau\!\!\int_0^1\!\!\d t
\left(\!\frac{-x^2}{4}\!\right)^{\!k}\!\!
\frac{\square^k}{k!k!}
\left(\!\frac{1-t}{t}\!\right)^{\!k}\!\!
O_\mu(0,\kappa\tau t x) .
\nonumber
\end{align}
%\vspace{-.3cm}
For the harmonic vector operator of twist $\tau = 3$ we obtain:
\begin{eqnarray}
\label{tw3_ir}
O^{\mathrm{tw3}}_{\beta} (0,\kappa x)
&=&
\left(\delta_{\beta}^{\alpha}(x\pd)-x^\alpha\pd_\beta\right)
\int_{0}^{1} \d\lambda\,
\tl O_\alpha (0,\kappa\lambda x),
\end{eqnarray}
and its generalization 
to arbitrary values of $(\ka, \kb)$ simply reads
\begin{eqnarray}
\label{tw3_irk}
O^{\mathrm{tw3}}_{\beta} (\ka x,\kb x)
&=&
\left(\delta_{\beta}^{\alpha}(x\pd)-x^\alpha\pd_\beta\right)
\int_{0}^{1} \d\lambda\,
\tl O_\alpha (\ka\lambda x,\kb\lambda x);
\end{eqnarray}
it fulfills the equations
\begin{eqnarray}
\square O^{\mathrm{tw3}}_{\beta}  (\ka x,\kb x)=0,
\qquad
\pd^\beta O^{\mathrm{tw3}}_{\beta}  (\ka x,\kb x)=0\ .
\end{eqnarray}
Because of the various differential operators which come into
the play by formulas (\ref{tw3_ir}) or (\ref{tw3_irk}) it is
not as easy as in the case of scalar operators to discriminate
between the different trace terms.

In the limit $x \rightarrow \xx$ only the terms with
 $k=0, 1$ and $2$ of (\ref{proj_tw3})
survive. The integrals which appear
in these formulas may be rewritten as:
\begin{eqnarray}
\int_0^1\!\!\!\d\lambda\int_0^1\!\!\frac{\d t}{t} 
f(\kappa\lambda t)
&=&
\int_0^1\!\!\d \lambda\,
\frac{1-\lambda}{\lambda} f(\kappa \lambda),
\\
\int_0^1\!\!\d\lambda\int_0^1\!\!\d\tau\int_0^1\!\!\d t 
f(\kappa\lambda\tau t)
&=&
\int_0^1\!\!\d \lambda 
\big(\hbox{\large$\frac{1}{2}$}
\ln^2 \!\lambda\big) f(\kappa\lambda),
\\
\int_0^1\!\!\d\lambda\int_0^1\!\!\d\tau\int_0^1\!\!\d t
\frac{1-t}{t}
 f(\kappa\lambda\tau t)
&=&
\int_0^1\!\!\d \lambda\Big(
\frac{1-\lambda}{\lambda}
+\ln \lambda -
\hbox{\large$\frac{1}{2}$}
\ln^2\! \lambda\Big) f(\kappa \lambda) .\qquad
\end{eqnarray}
Then for the nonlocal light--ray operator of twist--3 we get:
\begin{eqnarray}
\lefteqn{
O^{\mathrm{tw3}}_{\beta} (0,\kappa\tilde{x})
=
\int_{0}^{1}\d\lambda
\Big(g_{\beta\alpha}(x\pd)-x_\alpha\pd_\beta\Big)
\bigg[\!\!
\left\{1-\hbox{\large$\frac{1-\lambda}{\lambda}$}\,
\hbox{\large$\frac{x^2}{4}$}
\square\right\}g^{\alpha\mu}}
%\nonumber
\\
& &
-\big[ x^\alpha\pd^\mu(x\pd)
-\hbox{\large$\frac{1}{2}$}
x^2\pd^\alpha\pd^\mu\big]\!\!
%\nonumber\\
%& &
%\times
\left(\hbox{\large$\frac{\ln^2\lambda}{2}$}
-\left\{\!\hbox{\large$\frac{1-\lambda}{\lambda}$}
+\ln\lambda-\hbox{\large$\frac{\ln^2\lambda}{2}$}\right\}
\!\hbox{\large$\frac{x^2}{4}$}\square\right)\!\!\bigg]
O_\mu(0,\kappa\lambda x)\Big|_{x=\tilde{x}}.
\nonumber
\end{eqnarray}

Now, the various differentiations should be carried out. 
Again, using
$(x\pd)f(\lambda x)\\=\lambda\pd f(\lambda x)/\pd\lambda $ 
and performing the partial integrations with respect to $\lambda$,
after generalization to arbitrary values of 
$\kappa_i$, we finally obtain the simple expression:
\begin{align}
\label{O3k}
%\lefteqn{
O^{\mathrm{tw3}}_{\beta} (\ka\xx,\kb\xx)
=&
\int_{0}^{1}\d\lambda
\Big[(\delta_{\beta}^{\alpha}(x\pd)
  -x^\alpha\pd_\beta)\\
&\;  
-x_\beta\Big\{(\ln\lambda)
x^\alpha\square
 +(1+2\ln\lambda)\pd^\alpha\Big\}\Big]
O_\alpha(\kappa_1\lambda x,\kappa_2\lambda x)
\Big|_{x=\tilde{x}}.
\nonumber
\end{align}
In axial gauge we obtain (with the above mentioned conventions)
\begin{align}
\hspace{-.3cm}
O^{\mathrm{tw3}}_{\beta} (\ka\xx,\kb\xx)
=
\int_{0}^{1}\!\! \d\lambda
\bar{\psi}(\ka y)\Big(&
\big[\gamma_\beta(y {\sf D})-\lcyg {\sf D}_\beta\big]
%\nonumber
\\
& -y_\beta\left[(\ln\lambda)\lcyg
{\sf D}^2+(1+2\ln\lambda){\sf D}\slass\right]\!\!\Big)\psi(\kb y)
\Big|_{y=\lambda\tilde{x}} .
\nonumber
\end{align}

Again, the trace terms of (\ref{O3k})
being proportional to $x_\beta$
determine the (partially) antisymmetric bilocal twist--4 vector 
operator:
\bea
\hspace{-.5cm}
\label{tw4_as}
O^{\mathrm{tw4}}_{\beta,{\rm as}} (\ka\xx,\kb\xx)
=
\xx_\beta\!\!
\int_{0}^{1}\!\!\d\lambda
\big\{(\ln\lambda)
x^\alpha\square
 +(1+2\ln\lambda)\pd^\alpha\big\}
O_\alpha(\kappa_1\lambda x,\kappa_2\lambda x)
\Big|_{x=\tilde{x}}.
\eea

Now, having determined the nonlocal operators of twist 2 and 3 
we can use either eq.~(\ref{O_tw_nl}) 
to get the nonlocal light--cone operator
of twist $\tau=4$ which includes the contributions of the Young
patterns (i) and (ii) with spin $n-1$ being contained
in ${\bf T}(\frac{n-1}{2},\frac{n-1}{2})$, or simply add 
both twist--4 contributions:
\begin{align}
\hspace{-.5cm}
\label{O_tw4}
O^{\mathrm{tw4}}_{\beta} (\kappa_1\xx,\kappa_2\xx)
&=
O^{\mathrm{tw4}}_{\beta,{\rm sym}} (\kappa_1\xx,\kappa_2\xx)
+
O^{\mathrm{tw4}}_{\beta,{\rm as}} (\kappa_1\xx,\kappa_2\xx)
\nonumber\\
&=
\xx_\beta \int_{0}^{1}\d\lambda
\Big[(1+\ln\lambda) \pd^\alpha+
\hbox{\large$\frac{\ln\lambda}{2}$}
 x^\alpha\square\Big]
O_\alpha(\kappa_1\lambda x,\kappa_2\lambda x)
\Big|_{x=\tilde{x}};
%\nonumber
\end{align}
and for the axial gauge we may write
\begin{eqnarray}
\hspace{-.3cm}
O^{\mathrm{tw4}}_{\beta} (\kappa_1\xx,\kappa_2\xx)
&=&
\xx_\beta\int_{0}^{1} \d\lambda\lambda\,
\bar{\psi}(\ka y)\Big((1+\ln\lambda){\sf D}\slass
+\hbox{\large$\frac{\ln\lambda}{2}$}\lcyg
{\sf D}^2\Big)\psi(\kb y)\Big|_{y=\lambda\xx}\!.\qquad
\end{eqnarray}
The corresponding local twist--4 operators read:
\begin{equation}
O^{\rm tw4}_{\beta n}
=
\frac{\xx_\beta}{2(n+1)^2}\big(\pd^\alpha (x\pd)
- x^\alpha\square\big)
\bar{\psi}(0)\gamma_\alpha (x D)^n\psi(0)\Big|_{x=\xx}\, .
\end{equation}

This finishes the decomposition of the nonlocal vector quark
operators into its components of definite twist. 
Quite analogous results obtain for the axial vector operator by
simply replacing $\gamma_\beta$ by $\gamma_\beta\gamma_5$.

%\end{document}
%
%%%%%%%%%%%%%%%%%%%%%%%%%%%%%%%%%%%%%%%
%%%%%
\subsection{Antisymmetric tensor operators}
\label{tensor}
%%%%%%%%%%%%%%%%%%%%%%%%%%%%%%%%%%%%%%%
%%%%%
%
%
Now, together with its (partly) contracted operator
$M_{\alpha}= x^\beta M_{\alpha\beta}$, 
we consider the following tensor operator~\footnote{
The contracted form of the antisymmetric tensor operator
appears in applications, 
e.g.,~to the Drell--Yan process (see~\cite{Jaf92,Jaf96}).
Sometimes, especially in connection with expressions
like $(\xx\gamma) (i\gamma D)$, also the (local) tensor operator 
with $\ii \gamma_\alpha \gamma_\beta = \sigma_{\alpha\beta} 
+ \ii g_{\alpha\beta}$ has been used \cite{Kod79,Gey96b} 
which, however, is not 
irreducible. We denoted these tensors by $M$ since they are
related to the twist--3 contribution of the
quark mass in polarized deep inelastic scattering \cite{Shu82,Gey96b}. 
} 
\bea
\label{Mab}
M_{\alpha\beta}(0,\kappa x)
&=&
\bar{\psi}(0)\sigma_{\alpha\beta}
U(0,\kappa x)\psi(\kappa x),\\
\label{Ma}
M_{\alpha}(0,\kappa x)
&=&
\bar{\psi}(0)\sigma_{\alpha\beta}x^\beta
U(0,\kappa x)\psi(\kappa x) .
\eea
%with $\Gamma$ being given by 
%$\sigma_{\mu\nu} 
%=(i/2){\epsilon_{\mu\nu}}^{\kappa\lambda}\gamma_5
%\sigma_{\kappa\lambda}$.
The corresponding local tensor operators are given by 
\begin{eqnarray}
\label{M_loc}
\hspace{-1cm}
M_{\alpha\beta\mu_1\ldots\mu_n}
\!\!\!\!&\equiv&\!\!\!
\bar{\psi}(0)\sigma_{\alpha\beta}D_{\{\mu_1}\ldots
D_{\mu_n\} } \psi (0),
\\
\hspace{-1cm}
\label{M_loc1}
\!\!\!\!&=&\!\!\!
\alpha_{n+1}
\bar{\psi}(0)\sigma_{\alpha\{\beta}D_{\mu_1}\ldots
D_{\mu_n\} } \psi (0)
\!+\!
\beta_n
\bar{\psi}(0)\sigma_{[\alpha\beta}D_{\{\mu_1]}\ldots
D_{\mu_n\} } \psi (0)
\\
\hspace{-1cm}
\!\!\!\!&&\!\!\! + \ldots\ ,
\nonumber
\end{eqnarray}
which also decompose into two parts being related to the Young 
patterns (ii) and (iii) and further contributions being partially
antisymmetric with respect to the $\mu_i$'s;
$\alpha_{n+1}= 2(n+1)/(n+2)$ is already known,
and $\beta_n = 3n/(n+2)$ is the normalizing factor of 
${\cal Y}_{[m]}$ with $[m] = (n,1,1)$.
These operators have mass dimension $n+3$; for $n=0$ the operator
 transforms as an antisymmetric tensor of rank 2 and
therefore has spin $j=1$, twist $\tau = 2$. For $n > 0$
these tensors are reducible. The corresponding Clebsch--Gordan
series is given by
\begin{eqnarray}
\label{CGR2}
\lefteqn{
\big((1,0)\oplus (0,1)\big)\otimes
\left(
\hbox{\large$\left(\frac{n}{2},\frac{n}{2}\right)$}\oplus
\hbox{\large$\left(\frac{n-2}{2},\frac{n-2}{2}\right)$}\oplus
\ldots\right)}
\\
&=&
\left(
\hbox{\large$\left(\frac{n+2}{2},\frac{n}{2}\right)$}\oplus
\hbox{\large$\left(\frac{n}{2},\frac{n+2}{2}\right)$}
\right)\oplus 2
\hbox{\large$\left(\frac{n}{2},\frac{n}{2}\right)$}\oplus 2
\left(
\hbox{\large$\left(\frac{n-2}{2},\frac{n}{2}\right)$}\oplus
\hbox{\large$\left(\frac{n}{2},\frac{n-2}{2}\right)$}
\right)\oplus\ldots\ .\quad
\nonumber
\end{eqnarray}
Again, as in the case of the vector operators, 
eq.~(\ref{O_tw}), this corresponds --
for any fixed value of $n$ -- to a finite twist decomposition
$(\tau \geq 2)$. In the limit $x \rightarrow \xx$ only the
first three (respectively two) terms survive to get
\begin{eqnarray}
\label{M_tw}
 M_{\alpha\beta}(0,\kappa \xx)
&=&
 M^{\mathrm{tw2}}_{\alpha\beta}(0,\kappa \xx)
+M^{\mathrm{tw3}}_{\alpha\beta}(0,\kappa \xx)
+M^{\mathrm{tw4}}_{\alpha\beta}(0,\kappa \xx)\ ,
\\
\label{M_tw1}
 M_{\alpha}(0,\kappa \xx)
&=&
 M^{\mathrm{tw2}}_{\alpha}(0,\kappa \xx)
+M^{\mathrm{tw3}}_{\alpha}(0,\kappa \xx)\ .
\end{eqnarray}
The local tensors of symmetry class (ii) -- the only ones being 
relevant for $M_{\alpha}(0,\kappa \xx)$ -- contribute to $\tau = 2,3$,
whereas those of symmetry class (iii) contribute to
$\tau = 3,4$. This may be seen as follows:
In the case (ii) 
the trace terms being proportional to 
%$\xx_\beta \sum_i g_{\alpha\mu_i}$ and 
$\xx_{[\alpha}\sum_i g_{\beta] \mu_i}$ correspond to spin $j=n, n-1$ 
and, therefore, to twist $\tau = 3,4$, and those being proportional 
to $x^2 g_{\mu_i \mu_j}$ 
correspond to spin $j=n-1$ and twist $\tau = 4$, but 
they disappear on the light--cone.
In the case of symmetry class (iii) some of the latter 
contributions remain due 
to an additional derivation acting on $x^2$ 
(see eq.~(\ref{M_tw3_nl_ir})). From this consideration 
it becomes obvious that the twist--2 and twist--4 contributions 
of $M_{\alpha\beta}(0, \ka\xx)$ are uniquely defined
on the light--cone, whereas the twist--3 contributions consist of
two parts.

\smallskip
\noindent
(A)~~~{\em Tensor operators of symmetry class (ii) and their
contractions}\\
Now we consider tensor operators (and their contractions with $x$)
having symmetry class (ii) which is determined
by the following standard tableaux:
\\
\\
\unitlength0.5cm
\begin{picture}(30,1)
\linethickness{0.075mm}
\put(1,-1){\framebox(1,1){$\alpha$}}
\put(1,0){\framebox(1,1){$\beta$}}
\put(2,0){\framebox(1,1){$\mu_1$}}
\put(3,0){\framebox(1,1){$\mu_2$}}
\put(4,0){\framebox(3,1){$\ldots$}}
\put(7,0){\framebox(1,1){$\mu_n$}}
\put(8.5,0){$\stackrel{\wedge}{=}$}
\put(9.5,0)
{${\hbox{\large$\frac{2(n+1)}{n+2}$}}\!
\relstack{\alpha\beta}{\cal A}\;
\relstack{\beta\mu_1\ldots\mu_n}{\cal S} \!\!\!
\bar{\psi}(0)\sigma_{\alpha\beta}D_{\mu_1}\ldots
D_{\mu_n}\psi(0)\! - \!\mathrm{trace~terms,} $}
\end{picture}
\\ \\
with normalizing factor $\alpha_{n+1}$.
The corresponding twist--2 part transforms according to
%the (reducible w.r.~to $SO(1,3)$) representation
${\bf T}\hbox{$(\frac{n+2}{2},\frac{n}{2})\oplus
{\bf T}(\frac{n}{2},\frac{n+2}{2})$}$; it is given by:
\begin{align}
%\hspace{-1cm}
%\lefteqn{
M^{\mathrm{tw2}}_{\alpha\beta\mu_1\ldots\mu_n}
\label{M_tw2_l}
=&\;
\hbox{\large$\frac{1}{n+2}$}
\Big(
2\bar{\psi}(0) \sigma_{\alpha\beta} D_{\{\mu_1}\ldots
D_{\mu_n\} } \psi (0) %}
\\
& 
+\sum\limits_{l=1}^{n}
\bar{\psi}(0) \sigma_{\alpha\{\mu_l}
D_{\mu_1}\ldots D_{\mu_{l-1}}D_{|\beta|} D_{\mu_{l+1}}
\ldots D_{\mu_n\} } \psi(0)
\nonumber\\
& 
-\sum\limits_{l=1}^{n}
\bar{\psi}(0) \sigma_{\beta\{\mu_l}
D_{\mu_1}\ldots D_{\mu_{l-1}}D_{|\alpha|}
D_{\mu_{l+1}}\ldots D_{\mu_n\} } \psi(0)\Big)
-\mathrm{trace~terms}\, .%\qquad
\nonumber
\end{align}
%The prefactor has been choosen as in eq.~(\ref{prefac}) for
%the symmetry class (ii), but with $n+1$ instead of $n$.
Of course, this representation is reducible with respect to
$SO(1,3)$, because in that case it is possible to discriminate between the
selfdual and antiselfdual part related to $\sigma^\pm_{\mu\nu}$.
However, in physical applications this is of no relevance,
since  only the orthochronous Lorentz group $O(1,3)$ is relevant
and then parity connects them.
After contraction with $x^{\mu_1}\ldots x^{\mu_n}$ we obtain:
\begin{eqnarray}
\label{Mx_tw2_l}
M^{\mathrm{tw2}}_{\alpha\beta n}(x)
%=
%x^{\mu_1}\ldots x^{\mu_n}
%M^{\mathrm{tw2}}_{\alpha\beta\mu_1\ldots\mu_n}
&=&
\hbox{\large$\frac{2}{n+2}$}
\pd_{[\beta}
{\bar\psi}(0)\sigma_{\alpha]\rho}x^{\rho}(x D)^n\psi(0)
-\mathrm{trace~terms},
\\%\qquad
%\end{eqnarray}
%and 
%\begin{eqnarray}
M^{\mathrm{tw2}}_{\alpha n+1}(x)
&=&
{\bar\psi}(0)\sigma_{\alpha\rho}x^{\rho}(x D)^n\psi(0)
-\mathrm{trace~terms},
\end{eqnarray}
respectively.\footnote{
Here, the same comments are in order as in 
footnote \ref{SYM}.
Let us remind that according to the definition of the 
antisymmetrization it holds 
$2 T_{[\mu\nu]} = T_{\mu\nu} - T_{\nu\mu}$.
%Introducing an additional derivation $D_\alpha$ into these expressions
%leads to operators of twist--3.
}
Proceeding in the same way as in the preceeding Subsections
 we sum up to obtain the nonlocal twist--2 operators:
\begin{align}
\hspace{-1cm}
\label{M_tw2_nl}
M^{\mathrm{tw2}}_{\alpha\beta} (0,\kappa x)
&=
2 \int_{0}^{1} \d\lambda\,\lambda\,\pd_{[\beta}
\Big(\!
M_{\alpha]}(0,\kappa \lambda x)
-\mathrm{trace~terms}\Big),\\%\qquad
%\end{eqnarray}
%and
%\begin{eqnarray}
\hspace{-1cm}
\label{M2}
M^{\mathrm{tw2}}_{\alpha} (0,\kappa x)
&=
\Big(\delta_{\alpha}^{\mu}(x\pd)-x^\mu\pd_\alpha\Big)\!
\int_{0}^{1} \!\d\lambda\lambda 
\Big(\!
M_{\mu}(0,\kappa\lambda x)-\mathrm{trace~terms}\Big),
%\qquad
\end{align}
respectively. 
Here, we remark that, due to the extra factor $x^\beta$
in eq.~(\ref{Ma}), also
an extra factor $\lambda$ appears in the integrand of 
eqs.~(\ref{M_tw2_nl}), (\ref{M2}) and, furthermore,
 that eq.~(\ref{M2}) has exactly the
same structure as in the case of vector operators.

Taking into account eq.~(\ref{Proj1}) for the local traceless 
tensor $\tl M_{\alpha n+1}$, the integral representations of
$1/(n+2)$ and of the beta function we arrive at the following
expression for the nonlocal traceless (vector) operator
\begin{align}
\tl M_\alpha(0,\kappa x)
=&\,
M_\alpha(0,\kappa x)
+\sum_{k=1}^\infty\int_0^1\!\d t
\left(\frac{-x^2}{4}\right)^{\!k}\!
\frac{\square^k}{k!(k-1)!}
\left(\frac{1-t}{t}\right)^{\!k-1}\!\!
M_\alpha(0,\kappa t x)
\nonumber\\
\label{M2_tl}
& 
- \big[ x_\alpha\pd^\mu(x\pd)
%-x_\alpha x_\mu\square
- \hbox{\large$\frac{1}{2}$} 
x^2\pd_\alpha\pd^\mu\big]
\\
& 
\times
\sum_{k=0}^\infty
\int_0^1\!\d\tau\tau \int_0^1\!\d t\,t
\left(\frac{-x^2}{4}\right)^{\!k}
\frac{\square^k}{k!k!}
\left(\frac{1-t}{t}\right)^{\!k}\!
M_\mu(0,\kappa\tau t x)\ .
\nonumber
\end{align}
From this the irreducible twist--2 operators are obtained as follows:
\begin{eqnarray}
\label{Mab2}
M^{\mathrm{tw2}}_{\alpha\beta}(0,\kappa x)
&=&
2 \int_{0}^{1} \d\lambda\lambda\,
\pd_{[\beta}\tl M_{\alpha]}(0,\kappa\lambda x),
\\
\label{Ma2}
M_\alpha^{\rm tw2}(0,\kappa x)
&=&
\tl M_\alpha(0,\kappa x).
\end{eqnarray}
To obtain the last equation use has been made of a partial integration; 
%and $x_\mu\pd_\nu = \pd_\nu x_\mu - g_{\mu\nu}$; it 
the result is
in conformity with the twist--2 case of the vector operators
(compare eq.~(\ref{O_tw2}) and the related remarks).
Again, these operators are harmonic tensor functions:
\begin{align}
\label{M2harm}
\square M^{\mathrm{tw2}}_{\alpha\beta}(0,\kappa x) = 0,
\quad
\pd^\alpha M^{\mathrm{tw2}}_{\alpha\beta}(&0,\kappa x) =0,
\quad
\pd^\beta M^{\mathrm{tw2}}_{\alpha\beta}(0,\kappa x) =0
\\
\square M_\alpha^{\rm tw2}(0,\kappa x) =0,&
\quad
\pd^\alpha M_\alpha^{\rm tw2}(0,\kappa x)=0;
\end{align}
the first set of equations, 
due to (\ref{Mab2}), is a consequence of the second ones.

For the projection onto the light--cone only the terms 
in eq.~(\ref{M2_tl}) with $k=1,2$ contribute. 
Now, observing the following equalities,
\begin{align}
\hspace{-.6cm}
\int_0^1\!\!\d\lambda\lambda\!\int_0^1\!\!\d t 
f(\kappa\lambda t)
&=
\int_0^1\!\!\d\lambda
\left( 1-\lambda\right) f(\kappa\lambda),
\\
\hspace{-.6cm}
\int_0^1\!\!\d\lambda\lambda\!\int_0^1\!\!\d\tau\tau\!
\int_0^1\!\!\d t\, t
f(\kappa\lambda\tau t)
&=
\hbox{\large$\frac{1}{2}$}
\int_0^1\!\!\d\lambda\, 
\left( \lambda\ln^2 \lambda\right) f(\kappa\lambda),
\\
\hspace{-.6cm}
\int_0^1\!\!\d\lambda\lambda\!\int_0^1\!\!\d\tau\tau\!
\int_0^1\!\!\d t\left(1-t\right) f(\kappa\lambda\tau t)
&=\!
\int_0^1\!\!\d\lambda\Big(\!
1\!-\!\lambda\!+\!\lambda\ln \lambda\!-\!
\hbox{\large$\frac{\lambda}{2}$}
\ln^2\! \lambda\Big)
f(\kappa \lambda),\!\!\!\!
\end{align}
we obtain for the nonlocal twist--2 light--cone  tensor operator
\begin{align}
\hspace{-1cm}
%\lefteqn{
M^{\mathrm{tw2}}_{\alpha\beta} (0,\kappa\tilde{x})
=&
2 \int_{0}^{1}\d\lambda\lambda
\Big\{\delta_{[\alpha}^\mu\left(\pd_{\beta]}
-\hbox{\large$\frac{1-\lambda}{2\lambda}$}
x_{\beta]}\square\right)
-\hbox{\large$\frac{\ln^2\lambda}{2}$}
x_{[\alpha}\pd_{\beta]}\pd^\mu(1+x\pd)
%}
\\
&
+\hbox{\large$\frac{1}{2}$}
\left(\hbox{\large$\frac{1-\lambda}{\lambda}$}
+\ln\lambda\right)
\left(x_{[\alpha}\delta_{\beta]}^\mu
+x^\mu x_{[\alpha}\pd_{\beta]}\right)\square
\Big\}
M_\mu(0,\kappa\lambda x)\Big|_{x=\xx}\, .%\qquad
\nonumber
\end{align}
Performing the partial integrations which originate from 
$(x\pd) f(\lambda x) = \lambda \pd f/\pd\lambda$,
and generalizing to arbitrary values of $\kappa_i$,
we finally obtain
\begin{eqnarray}
\label{M_tw2_ir}
\hspace{-.6cm}
M^{\mathrm{tw2}}_{\alpha\beta} (\kappa_1\xx,\kappa_2\xx)
\!\!\!&=&\!\!\!
2 \int_{0}^{1}\!\!\d\lambda\,\lambda\,
\pd_{[\beta}
M_{\alpha]}(\kappa_1\lambda x,\kappa_2\lambda x)
\Big|_{x=\tilde{x}}
-
M^{\mathrm{higher}}_{\alpha\beta} (\kappa_1\xx,\kappa_2\xx),
\\
\label{M_hi.tw_a}
\hspace{-.6cm}
M^{\mathrm{higher}}_{\alpha\beta} (\kappa_1\xx,\kappa_2\xx)
\!\!\!&=&\!\!\!
 \int_{0}^{1}\!\!\d\lambda(1-\lambda)\!
\left(\!2 x_{[\alpha}\pd_{\beta]}\pd^\mu
-%\hbox{\large$\frac{1}{2}$}
x_{[\alpha}\delta_{\beta]}^\mu\square\!
\right)\!
M_\mu(\kappa_1\lambda x,\kappa_2\lambda x)
\Big|_{x=\tilde{x}} .
\nonumber
\end{eqnarray}
This last expression contains both twist--3 and twist--4 contributions.
%%, for later use, will be rewritten as follows
After an explicit decomposition of 
$M^{\mathrm{higher}}_{\alpha\beta}$ we get
\begin{equation}
\label{M_tw2_b}
M^{\mathrm{higher}}_{\alpha\beta}
(\kappa_1\tilde{x},\kappa_2\tilde{x})
=
M^{\mathrm{tw3}}_{\alpha\beta, a}(\kappa_1\tilde{x},\kappa_2\tilde{x})+
M^{\mathrm{tw4}}_{\alpha\beta, a}(\kappa_1\tilde{x},\kappa_2\tilde{x})
\end{equation}
with
\begin{eqnarray}
\label{M_tw3_ii}
M^{\mathrm{tw3}}_{\alpha\beta, a}
(\kappa_1\tilde{x},\kappa_2\tilde{x})
\!\!\!&=&\!\!\!
- \int_{0}^{1}\d\lambda\,
\hbox{\large$\frac{1-\lambda^2}{\lambda}$}
x_{[\alpha}\pd_{\beta]}x^{[\mu}\pd^{\nu]}
M_{\mu\nu}(\kappa_1\lambda x,\kappa_2\lambda x)
\Big|_{x=\xx}\\
M^{\mathrm{tw4}}_{\alpha\beta, a}
(\kappa_1\tilde{x},\kappa_2\tilde{x})
\!\!\!&=&\!\!\!
\label{M_tw4_ii}
- \int_{0}^{1}\d\lambda \Big\{
\hbox{\large$\frac{(1-\lambda)^2}{\lambda}$}
x_{[\alpha}\big(\delta_{\beta]}^{[\mu}(x\pd)
-x^{[\mu}\pd_{\beta]}\big)\pd^{\nu]}
\nonumber\\
& & +(1-\lambda)x_{[\alpha}\delta_{\beta]}^{[\mu}x^{\nu]}\square\Big\}
M_{\mu\nu}(\kappa_1\lambda x,\kappa_2\lambda x)
\Big|_{x=\xx}.% \qquad
\end{eqnarray}
The twist--4 operator in eq.~(\ref{M_tw4_ii}) 
consists of two several twist--4 operators
with different prefactor.

Finally, performing the partial derivatives we obtain, 
especially for axial gauge,
\begin{eqnarray}
\lefteqn{
M^{\mathrm{tw2}}_{\alpha\beta}(\kappa_1\xx,\kappa_2\xx)
=
2 \int_{0}^{1}\d\lambda\lambda
\bar{\psi}(\ka y)\Big\{
\left[\sigma_{\alpha\beta}
-y^\rho\sigma_{\rho[\alpha}{\sf D}_{\beta]}\right]
}\\
& &
+\hbox{\large$\frac{1-\lambda}{\lambda}$}
\Big[
%y_{[\alpha}\sigma_{\beta]\rho}{\sf D}^\rho+
2y_{[\alpha}\sigma_{\beta]\mu}{\sf D}^\mu
+y^\rho \sigma_{\rho\mu} y_{[\alpha}{\sf D}_{\beta]}{\sf D}^\mu
+ \hbox{\large$\frac{1}{2}$}
 y_{[\alpha}\sigma_{\beta]\rho}y^\rho{\sf D}^2
\Big]\Big\}\psi(\kb y)\Big|_{y=\lambda\xx} .
\nonumber
\end{eqnarray}

For the partially contracted nonlocal twist--2 light--cone operator
we obtain
\begin{align}
M^{\mathrm{tw2}}_{\alpha}(0,\kappa\tilde{x})
=
M_\alpha(0,\kappa x)
%\nonumber\\
+
\xx_\alpha
\Big(\pd_\mu(x\pd)
- \hbox{\large$\frac{1}{2}$} x_\mu\square\Big)\!\!
\int_0^1\d\lambda\left(\lambda\ln\lambda\right)
M^\mu(0,\kappa\lambda x)\Big|_{x=\tilde{x}},
\nonumber
\end{align}
from which, after partial integration and generalization
to arbitrary $\kappa_i$ we obtain~\footnote{
This result for $\ka = 1, \kb =0$, and including $x^2$--terms,
has been obtained earlier in Ref.~\cite{BBK89}} 
\begin{align}
\label{M_tw2_V}
M^{\mathrm{tw2}}_{\alpha}(\ka\xx,\kb\xx)
=&\;
{\bar\psi}(\kappa_1\xx)\sigma_{\alpha\rho}\xx^{\rho}
U(\kappa_1 \xx,\kappa_2\xx)\psi(\kappa_2\xx)
%\nonumber
\\
& 
+\xx_\alpha \xx^{\rho}\pd^\mu
\int_0^1\d\lambda\lambda
{\bar\psi}(\kappa_1\lambda x)\sigma_{\rho\mu}
U(\kappa_1 \lambda x,\kappa_2\lambda x)
\psi(\kappa_2\lambda x)\Big|_{x=\tilde{x}},\nonumber
\end{align}
and, correspondingly, in axial gauge
\begin{equation}
M^{\mathrm{tw2}}_\alpha(\ka\xx,\kb\xx)
=
%M_\alpha(\ka\xx,\kb\xx)
\bar{\psi}(\ka\xx)\sigma_{\alpha\rho}\xx^{\rho}\psi(\kb\xx)
%\nonumber\\
%& &
+\xx_\alpha\!\!\int_0^1\!\!\d\lambda%\lambda
\bar{\psi}(\ka y)y^{\rho}\sigma_{\rho\mu}{\sf D}^\mu
\psi(\kb y)\Big|_{y=\lambda \xx}.
%\nonumber
\end{equation}
Because the first term on the r.h.~side equals 
$M_\alpha(0,\kappa \xx)$ its twist--3 part according to 
eq.~(\ref{M_tw2_V}) is given by the second term 
(up to a minus sign):
\begin{eqnarray}
\hspace{-1cm}
M^{\mathrm{tw3}}_{\alpha}(\ka\xx,\kb\xx)
\!\!\!&=& \!\!\!
\xx_\alpha \xx^{\rho}\pd^\mu\!
\int_0^1\!\!\d\lambda\lambda
{\bar\psi}(\kappa_1\lambda x)\sigma_{\mu\rho}
U(\kappa_1 \lambda x,\kappa_2\lambda x)
\psi(\kappa_2\lambda x)\Big|_{x=\tilde{x}}.
\end{eqnarray}
Its local components,
which already have been used in~\cite{Jaf92,Koi95}, are
\begin{equation}
M^{\mathrm{tw3}}_{\alpha n+1}
=
\hbox{\large$\frac{1}{n+2}$}\xx_\alpha
\bar{\psi}(0)
\sigma_{\mu\rho}\xx^{\rho}(\xx D)^n D^\mu\psi(0)\ .
\end{equation}

%%%%%%%%%%%%%%%%%%%%%%%%%%%%%%%%
\smallskip
\noindent
(B)~~~{\em Tensor operators of symmetry class (iii)}\\
Now we consider the twist--3 and twist--4 contributions originating
from the (only possible) standard tableaux for the symmetry
class (iii):\\
%\pagebreak
\\
\unitlength0.5cm
\begin{picture}(30,1)
\linethickness{0.075mm}
\put(1,-2){\framebox(1,1){$\alpha$}}
\put(1,-1){\framebox(1,1){$\beta$}}
\put(1,0){\framebox(1,1){$\mu_1$}}
\put(2,0){\framebox(1,1){$\mu_2$}}
\put(3,0){\framebox(1,1){$\mu_3$}}
\put(4,0){\framebox(3,1){$\ldots$}}
\put(7,0){\framebox(1,1){$\mu_n$}}
\put(8.5,0){$\stackrel{\wedge}{=}$}
\put(9.5,0)
{$\hbox{\large$\frac{3n}{n+2}$}\!
\relstack{\alpha\beta\mu_1}{\cal A}\,
\relstack{\mu_1\ldots\mu_n}{\cal S}\!\!
\bar{\psi}(0)\sigma_{\alpha\beta}D_{\mu_1}\ldots
D_{\mu_n}\psi(0)\! -\mathrm{trace~terms.} $}
\end{picture}
\\ \\ \\
The corresponding traceless local tensor having twist $\tau = 3$ and being
contained in ${\bf T}\!\left(\frac{n}{2},\frac{n}{2}\right)$ is
given by (where $\beta_n = 3n/(n+2)$ has been choosen):
\begin{eqnarray}
\label{M3}
M^{\mathrm{tw3}}_{\alpha\beta\mu_1\ldots\mu_n}
%&=&\frac{3n}{n+2}
%\relstack{\alpha\beta\mu_1}{\cal A}\,
%\relstack{\mu_1\ldots\mu_n}{\cal S}
%\bar{\psi}(0)\sigma_{\alpha\beta} D_{\mu_1}\ldots
%D_{\mu_n} \psi (0) -\mathrm{trace~terms} \nonumber \\
&=&
\hbox{\large$\frac{n}{n+2}$}
\Big(
\bar{\psi}(0)\sigma_{\alpha\beta} D_{\{\mu_1}\ldots
D_{\mu_n\} }\psi(0)
-\bar{\psi}(0)\sigma_{\alpha\mu_1} D_{\{\beta}\ldots
D_{\mu_n\} } \psi(0)
\nonumber\\
\label{M_tw3_l}
& & \qquad
+\bar{\psi}(0)\sigma_{\beta\mu_1}
D_{\{\alpha}\ldots D_{\mu_n\} } \psi(0)\Big)
-\mathrm{trace~terms}.
\end{eqnarray}
Again, the prefactor $\beta_n$ could have been determined by the equality
(compare eq.~(\ref{prefac}))
\begin{equation}
\label{vorf}
M_{\alpha\beta n}(x)=
x^{\mu_1}\ldots x^{\mu_n}
\Big(
\hbox{\large$\frac{2(n+1)}{n+2}$}
\!\relstack{\alpha\beta}{\cal A}\,
\relstack{\beta\mu_1\ldots\mu_n}{\cal S}\!\!\!\!
M_{\alpha\beta\mu_1\ldots\mu_n}
+
\hbox{\large$\frac{3n}{n+2}$}
\!\relstack{\alpha\beta\mu_1}{\cal A}\,
\relstack{\mu_1\ldots\mu_n}{\cal S}\!\!\!
M_{\alpha\beta\mu_1\ldots\mu_n}\!\Big) .
\nonumber
\end{equation}
Now, contracting the expression (\ref{M3})  
with $x^{\mu_1}\ldots x^{\mu_n}$ we obtain:
\begin{eqnarray}
\label{Mx_tw3_l}
M^{\mathrm{tw3, b}}_{\alpha\beta n}(x)
&=&
x^{\mu_1}\ldots x^{\mu_n}
M^{\mathrm{tw3, b}}_{\alpha\beta\mu_1\ldots\mu_n}  \\
&=&
\hbox{\large$\frac{1}{n+2}$}
\left(%\hbox{\large$\frac{1}{2}$}
(x\pd)\delta^\rho_{[\beta}
- 2 x^\rho\pd_{[\beta}\right)
\Big(\bar{\psi}(0)\sigma_{\alpha]\rho}(x D)^n\psi(0)
%-(\alpha\leftrightarrow\beta)
-\mathrm{trace~terms}\Big) .
\nonumber
\end{eqnarray}
Resumming these local terms gives the nonlocal twist--3 operator 
as follows
\begin{eqnarray}
\label{M_tw3_nl_ir}
M^{\mathrm{tw3}}_{\alpha\beta, b}(0,\kappa x)
&=&
\int_0^1\d \lambda\lambda
\left(%\hbox{\large$\frac{1}{2}$}
(x\pd)\delta^\rho_{[\beta}
- 2 x^\rho\pd_{[\beta}\right)
\TLL {M_{\alpha]\rho}}(0,\kappa\lambda x)\, ,
\end{eqnarray}
where we introduced the traceless nonlocal operator
$\tl M_{\alpha\beta}(0,\kappa x)$ which results from
$\bar{\psi}(0)\sigma_{\alpha\beta}U(0,\kappa x)
\psi(\kappa x)$; it has to be determined by the
conditions 
\begin{equation}
\label{M0harm}
\square \tl M_{\alpha\beta n}=0\ ,
\qquad
\pd^\alpha \tl M_{\alpha\beta n}=0\ ,
\qquad
\pd^\beta \tl M_{\alpha\beta n}=0.
\end{equation}
The solution of these equations reads
(see Appendix \ref{trace}, eq.~(\ref{Proj5}))
\begin{align}
%\lefteqn{
\tl M_{\alpha\beta n}(x)
=&
\bigg\{\delta_\alpha^\mu\delta_\beta^\nu
-\hbox{\large$\frac{2}{n(n+1)(n+2)}$}
\bigg(x_{[\alpha}\pd_{\beta]}x^{[\mu}\pd^{\nu]}
%}
\\
&-
\left[
x_{[\alpha}\delta_{\beta]}^{[\mu}\pd^{\nu]}(x\pd)
-\hbox{\large$\frac{1}{2}$} x^2\pd_{[\alpha}
\delta_{\beta]}^{[\mu}\pd^{\nu]}\right](x\pd+2)\bigg)\bigg\}
%\quad\times
%\sum_{k=0}^{[\frac{n}{2}]}\frac{(n-k)!}{k!n!}\left
%(\frac{-x^2}{4}\right)^k \square^k 
H^{(4)}_n \!\big(x^2|\square\big)
M_{\mu\nu n}(x).
\nonumber
\end{align}
Using the integral representation of 
the additional factor $1/n$ (the remaining factors of the 
denominator are taken together with $1/n!$ to get $1/(n+2)!$) and of
the beta function we may sum up these local operators 
to the following expression:
\begin{eqnarray}
\lefteqn{\tl M_{\alpha\beta}(0,\kappa x)
=
M_{\alpha\beta}(0,\kappa x)
+\sum_{k=1}^\infty\int_0^1\!
\frac{\d t}{t}
\!\!\left(\!\frac{-x^2}{4}\right)^{\!k}\!\!
\frac{\square^k}{k!(k-1)!}
\left(\!\frac{1-t}{t}\right)^{\!k-1}\!\!
M_{\alpha\beta}(0,\kappa t x)}\qquad\qquad 
\nonumber\\
& &
-2 %\hbox{\large$\frac{1}{2}$}
\bigg\{
x_{[\alpha}\pd_{\beta]}x^{[\mu}\pd^{\nu]}
-\left[
x_{[\alpha}\delta_{\beta]}^{[\mu}\pd^{\nu]}(x\pd)
- \hbox{\large$\frac{1}{2}$} x^2\pd_{[\alpha}
\delta_{\beta]}^{[\mu}\pd^{\nu]}\right](x\pd+2)\bigg\}
\nonumber\\
& &\times\!\sum_{k=0}^\infty\!
\int_0^1\!\frac{\d\tau}{\tau}\!\int_0^1\!\!\d t\,t\!
\left(\!\frac{-x^2}{4}\right)^{\!\!k}\!\!
\frac{\square^k}{(k+1)!k!}
\left(\!\frac{1-t}{t}\right)^{\!\!k+1}
\!\! \!M_{\mu\nu}(0,\kappa\tau t x).\qquad\;
\end{eqnarray}
Then, by construction, eq.~(\ref{M_tw3_nl_ir}) defines 
a nonlocal operator of twist--3. As is easily seen by partial 
integration it fulfils the following relation
\begin{equation}
\label{M_tw3}
M^{\rm tw3}_{\alpha\beta, b}(0,\kappa x)
=\tl M_{\alpha\beta}(0,\kappa x)
- M^{\rm tw2}_{\alpha\beta}(0,\kappa x)\, 
\end{equation}
and, because of  eqs.~(\ref{M2harm}) and (\ref{M0harm})
it is a harmonic tensor operator.

Now, to determine the corresponding LC--operator 
we use  eq.~(\ref{M_tw3}).
Because $M^{\rm tw2}_{\alpha\beta}(0,\kappa \lcx)$ 
is already known we only have to determine 
$\tl M_{\alpha\beta}(0,\kappa\lcx)$ for which only the 
terms with $k=0,1$ are relevant.
Using the equalities
\begin{align}
\hspace{-.8cm}
\int_0^1\frac{\d\tau}{\tau}\int_0^1\d t (1-t)f(\kappa\tau t)
&=
\int_0^1\frac{\d t}{t}
\frac{\left( 1-t\right)^2}{2} f(\kappa t), 
\\
\hspace{-.8cm}
\int_0^1\frac{\d\tau}{\tau}\int_0^1\d t 
\frac{(1-t)^2}{t} f(\kappa\tau t)
&=
-\int_0^1\frac{\d t}{t}\Big(
\frac{\left(1-t\right)^2}{2}+(1-t)+\ln t\Big) \!f(\kappa t),
\end{align}
we finally obtain:
\begin{align}
\lefteqn{
\tl M_{\alpha\beta}(0,\kappa\lcx)
=  M_{\alpha\beta}(0,\kappa\lcx)
-%\hbox{\large$\frac{1}{4}$}\!\!
\left[ x_{[\alpha}\pd_{\beta]}x^{[\mu}\pd^{\nu]}
- x_{[\alpha}\delta_{\beta]}^{[\mu}\pd^{\nu]}
(x\pd)\big(x\pd+2\big)\!\right]}\qquad\\
&\times
\int_0^1\frac{\d \lambda}{\lambda}
\left((1-\lambda)^2 + \hbox{\large$\frac{1}{8}$}
\left\{(1-\lambda)^2+2(1-\lambda)+2\ln \lambda\right\}
x^2\square\right)
M_{\mu\nu}(0,\kappa\lambda x)\Big|_{x=\xx},%\qquad
\nonumber
\end{align}
which, after performing the partial integrations due to
$(x\pd)f(\lambda x)=\lambda\pd f(\lambda x)/\pd\lambda$,
leads to the expression:
\begin{align}
\label{M_tw3_a}
\tl M_{\alpha\beta}(0,\kappa\lcx)
=  M_{\alpha\beta}(0,\kappa\lcx)
&-%\hbox{\large$\frac{1}{4}$}
\int_0^1\frac{\d \lambda}{\lambda}
\Big\{(1-\lambda)^2
x_{[\alpha}\pd_{\beta]}x^{[\mu}\pd^{\nu]}\\
&\quad-
2\lambda x_{[\alpha}\delta_{\beta]}^{[\mu}\pd^{\nu]}
+(1-\lambda)x_{[\alpha}\delta_{\beta]}^{[\mu}x^{\nu]}
\square\Big\}
M_{\mu\nu}(0,\kappa\lambda x)_{|x=\tilde{x}}.\qquad
\nonumber
\end{align}

Now, inserting eqs.~(\ref{M_tw3_a}) and (\ref{M_tw2_b}) into
eq.~(\ref{M_tw3}) we obtain:
\begin{eqnarray}
\label{M_tw3_c}                                
\lefteqn{M^{\rm tw3}_{\alpha\beta, b}(0,\kappa \lcx)
=
M_{\alpha\beta}(0,\kappa\lcx)
-2 \int_{0}^{1}\d\lambda\lambda\,\pd_{[\beta}
M_{\alpha]}(0,\kappa\lambda x)\Big|_{x=\xx}}
\\
&& -\! \int_{0}^{1}\!\frac{\d\lambda}{\lambda}\!
\Big\{
(1\!-\!\lambda^2)\big(x_{[\alpha}\delta_{\beta]}^{[\mu}x^{\nu]}\square
+x_{[\alpha}\pd_{\beta]}x^{[\mu}\pd^{\nu]}\big)
%\nonumber\\&
\!-\!2\lambda^2 x_{[\alpha}
\delta_{\beta]}^{[\mu}\pd^{\nu]}\Big\}
M_{\mu\nu}(0,\kappa\lambda x)\Big|_{x=\xx}.
\nonumber
\end{eqnarray}
Let us write this for arbitrary $\kappa_i$ in the form
\begin{align}
\label{M_tw3_b}
M^{\rm tw3}_{\alpha\beta, b}(\ka\xx,\kb \xx)
=&
\int_{0}^{1}\d\lambda \lambda
\left(%\hbox{\large$\frac{1}{2}$}
(x\pd)\delta_{[\beta}^\nu
- 2 x^\nu\pd_{[\beta}\right)
M_{\alpha]\nu}(\kappa_1\lambda x,\kappa_2\lambda x)\Big|_{x=\xx}
\\
& 
- M^{\rm tw4}_{\alpha\beta, b}(\ka\xx,\kb \xx),
\nonumber
\end{align}
where the twist--4 part is determined by the trace
terms,
%of $M^{\rm tw3}_{\alpha\beta, b}(\ka\xx,\kb \xx)$, 
namely
\begin{align}
\label{M_tw4}
%\lefteqn{
%\hspace{-.5cm}
M^{\rm tw4}_{\alpha\beta, b}(\ka\xx,\kb \xx)
= 
&
- \int_{0}^{1}\d\lambda 
\hbox{\large$\frac{1-\lambda^2}{\lambda}$}\Big\{
x_{[\alpha}\big(\delta_{\beta]}^{[\mu}(x\pd)-x^{[\mu}\pd_{\beta]}\big)\pd^{\nu]}
\nonumber\\
&\ - x_{[\alpha}\delta_{\beta]}^{[\mu}x^{\nu]}\square\Big\}
M_{\mu\nu}(\kappa_1\lambda x,\kappa_2\lambda x)
\Big|_{x=\xx}.% \qqua
%%\nonumber
\end{align}

The full twist--4 operator gets from eqs.~(\ref{M_tw4_ii}) and (\ref{M_tw4})
\begin{eqnarray}
 \lefteqn{
M^{\rm tw4}_{\alpha\beta}(\ka\xx,\kb\xx)
=
M^{\rm tw4}_{\alpha\beta,a}(\ka\xx,\kb\xx)
+
M^{\rm tw4}_{\alpha\beta,b}(\ka\xx,\kb\xx)}
\\
&&=\!
 \int_{0}^{1}\d\lambda 
\hbox{\large$\frac{1-\lambda}{\lambda}$}\Big\{
x_{[\alpha}\delta_{\beta]}^{[\mu}x^{\nu]}\square
-2x_{[\alpha}\big(\delta_{\beta]}^{[\mu}(x\pd)-x^{[\mu}\pd_{\beta]}\big)\pd^{\nu]}
\Big\}M_{\mu\nu}(\kappa_1\lambda x,\kappa_2\lambda x)
\Big|_{x=\xx}.\nonumber
\end{eqnarray}

The complete twist-3 operator obtains from
eq.~(\ref{M_tw3_b}) and the trace terms with twist--3 of the twist--2
operator eq.~(\ref{M_tw3_ii}):
\begin{eqnarray}
%\label{M_tw3_e}
\lefteqn{M^{\rm tw3}_{\alpha\beta}(\ka\xx,\kb\xx)
=
M^{\rm tw3}_{\alpha\beta,a}(\ka\xx,\kb\xx)
+
M^{\rm tw3}_{\alpha\beta,b}(\ka\xx,\kb\xx)}
\nonumber\\
&&=\!
\int_{0}^{1}\!\!\d\lambda\Big\{
\lambda\left(%\hbox{\large$\frac{1}{2}$}
(x\pd)\delta_{[\beta}^\nu
- 2 x^\nu\pd_{[\beta}\right)\delta_{\alpha]}^{\mu}
-\hbox{\large$\frac{1-\lambda^2}{\lambda}$}\Big(
x_{[\alpha}\pd_{\beta]}x^{[\mu}\pd^{\nu]}
\\
& &\ +
x_{[\alpha}\big(\delta_{\beta]}^{[\mu}(x\pd)-x^{[\mu}\pd_{\beta]}\big)\pd^{\nu]}
- x_{[\alpha}\delta_{\beta]}^{[\mu}x^{\nu]}\square\Big)
\Big\}M_{\mu\nu}(\kappa_1\lambda x,\kappa_2\lambda x)
\Big|_{x=\xx}.
\nonumber
\end{eqnarray}
This completes the twist decomposition of the antisymmetric
tensor operators.

%%%%%%%%%%%%%%%%%%%%%%%%%
%%  Datei: lazgener.tex %
%%%%%%%%%%%%%%%%%%%%%%%%%
%\noindent

\section{Applications to distribution amplitudes and 
generalization to conformal LC operators}
\renewcommand{\theequation}{\thesection.\arabic{equation}}
\setcounter{equation}{0}
\label{general}

In this Section we like to point out one of the virtues of the
twist decomposition which have been given in Section
\ref{twist} by considering nonforward matrix elements
of the nonlocal LC operators with definite twist.
Furthermore, we introduce the generalization of the twist
decomposition to conformal light--cone operators.

%%%%%%%%%%%%%%%%%%%%%%%%%%%%%%%
      \subsection{Matrix elements of LC operators
     with definite twist}
%%%%%%%%%%%%%%%%%%%%%%%%%%%%%%%
%
As it became obvious there exist relations between nonlocal operators
of the same twist but different tensor structure, e.g., between the 
vector and scalar operators, 
cf.~Eqs.~(\ref{O2_betak}) and (\ref{O2k}) as well as 
Eqs.~(\ref{Mab2}) and (\ref{Ma2}),
and also of different twist and same tensor structure, 
cf.~Eqs.~(\ref{tw3_irk}) and (\ref{O2_betak}). This, of course, leads to
relations between the related distribution amplitudes 
obtained from the (non-forward) matrix elements.

The nonforward matrix elements of  light--ray operators
with definite twist between the hadron states $|p_i, S_i \rangle $ 
with momenta $p_i$ and spin $S_i, i = 1,2$, appear as 
nonperturbative inputs of various hard scattering processes.
Phenomenologically these quantities will be represented by
partition functions (in the forward case) resp.~distribution
amplitudes.
Because of translation invariance,
\begin{eqnarray}
\label{translation}
\langle p_2 \,| O^\Gamma(\ka x, \kb x) |\,p_1\rangle 
%&\equiv&
%\langle p_2 \,| O^\Gamma((\kappa_+ - \kappa_-) x,
% (\kappa_+ + \kappa_-) x) |\,p_1\rangle \nonumber \\
&=&
e^{i\kappa_+ xp_-} 
\langle p_2 \,|O^\Gamma(-\kappa_- x,\kappa_- x)|\,p_1\rangle,
\end{eqnarray}
it is completely sufficient to consider matrix elements
of the `centered' operators $O^\Gamma(-\kappa \xx,\kappa \xx)$.
Their Fourier transforms with respect to $\kappa\xx p_+$ and
$\kappa\xx p_-$ define the above mentioned distribution
amplitudes. 
%\\
In the following it is useful -- as it was the case for the twist 
decomposition -- to study the nonlocal operators temporarily 
for arbitrary values of $x$. Then, the distribution amplitudes 
on the light--cone are obtained as limiting case $x \rightarrow \xx$. 

First, we consider the scalar twist--2 quark operators
(\ref{proj_tw2}) 
between hadron states $|p_i, S_i \rangle$ and, in addition, we 
perform a kinematical decomposition of these matrix elements. 
If the equation of motion for the external hadron states are taken 
into account, then there remain only two independent
distribution amplitudes which may be choosen in the following
way ($u(p) \equiv u(p, s)$ is the free hadronic Dirac spinor and,
to simplify notations, $\sigma_{\mu\nu} x^\mu p_-^\nu $
is denoted by $(x \sigma p_-)$):
\begin{align}
\label{kdec1}
%\hspace{-.5cm}
\langle p_2|O^{\rm tw2}(-\kappa x,\kappa x)|p_1\rangle
& =   
\tilde g^{(2)}(\kappa xp_+,\kappa x p_-, 
\kappa^2 x^2, p_1p_2,\mu^2)
\; \bar u(p_2) (x \gamma)  u(p_1) \\
&+ \tilde h^{(2)}(\kappa x p_+, \kappa x p_-,
 \kappa^2 x^2, p_1p_2, \mu^2)
%\frac{1}{m}
\;M^{-1}\bar u(p_2)(x \sigma p_-) u(p_1).
\nonumber
\end{align}
Because of eq.~(\ref{O2_betak}) the twist--2 vector operator
is related to this kinematic decomposition:\footnote{
Here, it is necessary to remark that for the consideration
of anomalous dimensions and evolution equations of amplitudes 
it is mandatory to use the operators with general 
$\kappa$--values.}
\begin{eqnarray}
\label{kdecv}
\lefteqn{
\hspace{-.2cm}
\langle p_2|O^{\rm tw 2}_\mu
(-\kappa x, \kappa x)|p_1\rangle 
=
\int_0^1 \, d \lambda \,
\partial_\mu^x \,
\langle p_2|O^{\rm tw 2}(-\kappa\lambda x, \kappa\lambda x)
|p_1\rangle
} \nonumber \\
&= & 
\int_0^1 \, d \lambda \,
\partial_\mu^x \,
 \Big[\tilde g^{(2)}(\kappa \lambda x p_+,\kappa \lambda x p_-,
\kappa^2 \lambda^2 x^2, p_1p_2, \mu^2)
\; {\bar u}(p_2)(x \gamma) u(p_1)
\\
& & \qquad \quad +\;\; 
\tilde h^{(2)}(\kappa\lambda x p_+, \kappa\lambda x p_-,
 \kappa^2\lambda^2 x^2, p_1p_2, \mu^2)
\;M^{-1}\bar u(p_2)(x\sigma p_-)u(p_1)  \Big]. 
\nonumber
\end{eqnarray}
The application of this relationship between scalar and
vector operators to the virtual Compton scattering has 
been considered in detail in \cite{BGR98}. Especially,
it has been shown that any of the physical results about
the twist--2 vector operator is already contained in the
twist--2 scalar operator and its distribution amplitudes.
In order to indicate this roughly the Fourier transformations of 
$g$ and $h$ -- jointly being denoted by $f$ -- with respect to 
$\kappa xp_\pm$ as well as $\kappa^2 x^2$ are introduced,
\begin{eqnarray}
\label{zrep}
\tilde f(\kappa x p_+, \kappa x p_-, \kappa^2 x^2)
 =\! \int_{-1}^{+1}\!\!\! dz_+ dz_-
e^{-i\kappa \big((xp_+)z_+ + (xp_-)z_-\big)} 
\int\!\! dq  e^{-i\kappa xq}f(z_+,z_-, Q^2 ),
\nonumber
\end{eqnarray}
where unaffected arguments have been omitted.
Then the
$\lambda$--integration will be performed leading to
%--------------------------------------------------------------
\begin{eqnarray}
\label{MAIN}
\lefteqn{%\hspace{-.5cm}
\langle p_2|O^{\rm tw 2}_\mu (-\kappa x, \kappa
x)|p_1\rangle
 = \int_{-1}^{+1}\!\! dz_+ dz_-
e^{-i\kappa \big((xp_+)z_+ + (xp_-)z_-\big)} 
\!\!\int\!\! dq e^{-i\kappa xq}
} 
%\nonumber 
\\
& & \big\{ \, G(z_+,z_-, Q^2)\Big[
\delta_\mu^\lambda %\bar u(p_2)\gamma_\mu u(p_1) 
-i\kappa g_{\mu\nu}\left(p_+^\nu z_+ + p_-^\nu z_+ + q^\nu \right)
x^\lambda\big]\bar u(p_2) \gamma_\lambda u(p_1)
\nonumber \\
& &      
+ H(z_+,z_-, Q^2)\big[
\delta_\mu^\lambda %M^{-1}\bar u(p_2)(\sigma_{\mu \nu}p^\nu_-) u(p_1) 
-i\kappa g_{\mu\nu}\left(p_+^\nu z_+ + p_-^\nu z_+ + q^\nu \right)
x^\lambda\big]
M^{-1}\bar u(p_2)(\sigma_{\lambda \kappa}p^\kappa_-) u(p_1)  
\nonumber 
%\\& & \hspace{3cm}
%-i\kappa g_{\mu\nu}\left(p_+^\nu z_+ + p_-^\nu z_+ + q^\nu \right)
%M^{-1}\bar u(p_2)(x \sigma p_-) u(p_1) \Big] \Big\},  
%\nonumber
\end{eqnarray}
%---------------------------------------------------------------
where, again $F$ being either $G$ or $H$,
\begin{eqnarray}
\label{tt1}
F(z_+,z_-, Q^2) 
&=&
\int_0^1  \hbox{\large$\frac{d\lambda}{\lambda^6}$}
f\!\left(\hbox{\large$
\frac{z_+}{\lambda},\frac{z_-}{\lambda},
\frac{Q^2}{\lambda^2}$}\right)
\Theta (\lambda - |z_+|) \Theta (\lambda - |z_-|).
\nonumber
\end{eqnarray}
As has been proven in \cite{BGR98} if eq.~(\ref{MAIN})
is used in the nonlocal light--cone expansion of the 
product of the electromagnetic currents appearing in the
scattering amplitude of the virtual Compton scattering
the latter will be expressed by the double variable
distributions $G,H$, and the evolution of these amplitudes 
is governed by the anomalous dimensions of the twist--2
nonlocal {\em scalar} operators.

In the case of twist--3 vector operators the situation is more
complicated. However, the defining equation (\ref{tw3_irk}) 
may be rewritten somewhat and after partial integration
we obtain a relation which is similar to eq.~(\ref{M_tw3}):
\begin{eqnarray}
\label{tw3_irk1}
O^{\mathrm{tw3}}_{\beta} (-\kappa x,\kappa x)
&=&
\left(\delta_{\beta}^{\alpha}(x\pd + 1)-\pd_\beta x^\alpha\right)
\!\int_{0}^{1}\!\! \d\lambda\,
\tl O_\alpha (-\kappa\lambda x,\kappa\lambda x)
\nonumber
\\
&=&
\tl O_\beta (-\kappa x,\kappa x)
- 
O^{\mathrm{tw2}}_{\beta} (-\kappa x,\kappa x).
\end{eqnarray}
Furthermore, from eq.~(\ref{M_tw3}) we obtain the relation
\bea
x^\beta O^{\mathrm{tw3}}_{\beta} (-\kappa x,\kappa x) = 0.
\eea
Both equations restrict the nonforward matrix elements of
the twist--3 vector operators. From the last equation we get
\begin{align}
\label{k3dec1}
\hspace{-.5cm}
\langle p_2|O^{\rm tw3}_\mu(-\kappa x,\kappa x)|p_1\rangle
 = & \; 
(\delta^\alpha_\mu x^\beta - \delta^\beta_\mu x^\alpha)
\Big[
\tilde g^{(3)}_\alpha(\kappa xp_+,\kappa x p_-, 
\kappa^2 x^2%, p_1p_2,\mu^2
)
\; \bar u(p_2)  \gamma_\beta  u(p_1) 
\nonumber\\
&+ \tilde h^{(3)}_\alpha(\kappa x p_+, \kappa x p_-,
 \kappa^2 x^2%, p_1p_2, \mu^2
)
%(\delta^\alpha_\mu x^\beta - \delta^\beta_\mu x^\alpha)
\;M^{-1}\bar u(p_2) \sigma_{\beta\nu} p^\nu_-) u(p_1)
\Big]
\nonumber\\
&+{\tilde {h'}}^{(3)}_\alpha(\kappa x p_+, \kappa x p_-,
 \kappa^2 x^2%, p_1p_2, \mu^2
)
\bar u(p_2) \sigma_{\mu\nu}x^\nu u(p_1),
%\nonumber
\end{align}
where the functions $\tilde g^{(3)}_\mu,\tilde h^{(3)}_\mu$ 
and ${\tilde {h'}}^{(3)}_\mu$, because of eq.~(\ref{tw3_irk1}),
are related to $\tilde g^{(2)}$ and $\tilde h^{(2)}$ and the 
nonforward matrix elements of ${\tl O}_\alpha (-\kappa x,\kappa x)$; 
 however, the latter can not be determined so easily.

In the case of twist--2 tensor operators 
$M^{\mathrm{tw2}}_{\alpha\beta}(0,\kappa x)$
the situation is more simple since, because of 
eqs.~(\ref{Mab2}) and (\ref{Ma2}), we have the relation
\begin{eqnarray}
\label{Mab2k}
M^{\mathrm{tw2}}_{\alpha\beta}(-\kappa x,\kappa x)
&=&
2 \int_{0}^{1} \d\lambda\lambda\,
\pd_{[\beta} M_{\alpha]}^{\rm tw2}
(-\kappa\lambda x,\kappa\lambda x).
\end{eqnarray}
From this equation, being similar to (\ref{O2_betak}), 
relations may be derived which are analogous to 
eqs.~(\ref{kdec1}) and (\ref{kdecv}). Physical
consequences of this circumstance have not been
drawn up to now. The situation in the case of
twist--3 tensor operators is similar to the 
twist--3 vector operators above -- at least concerning
that part which obtains from symmetry type (ii).

%%%%%%%%%%%%%%%%%%%%%%%%%%%%%%%
      \subsection{ Conformal nonlocal LC operators
     with definite twist}
%%%%%%%%%%%%%%%%%%%%%%%%%%%%%%%
%
As is well known the diagonalization of the anomalous
dimensions which is essential for the solution of the
evolution equations of the distribution amplitudes
 -- at least in one-loop order -- is obtained by using
conformal light--cone operators. Therefore, the
results of this paper have to be generalized to conformal
operators. This, however, is easily achieved
since the light--cone operators 
$O_\Gamma(\ka\xx, \kb\xx)$ considered in Section 3 are 
related to the conformal ones by the following integral 
representation~\cite{BGHR85,BGHR87}:\footnote{
To be more precise, the operators 
$O_\Gamma^{\rm conf}(\kappa, t; \xx)$ obviously are scale
invariant, and the {\em local} conformal operators  
according to eqs.~(\ref{conloc1}) -- (\ref{conloc})
are directly related to them; see also the discussion of
eq.~(\ref{cloc}) below. Especially, they do not have 
definite conformal spin. However, the latter may be obtained 
by a more involved resummation of the local conformal 
operators which does not affect their twist decomposition.
 For a more detailed discussion
of nonlocal conformal operators see  Chapter 3 of 
Ref.~\cite{BB88}.}
\bea
\label{conform}
O_\Gamma^{\rm conf}(\kappa, t; \xx)
&=&
\pd_{(i\kappa)}\int \frac{d\rho}{2\pi}\,
O_\Gamma\big((\kappa - (1-t)\rho)\xx, (\kappa + (1+t)\rho)\xx
\big);
\\
\label{formcon}
O_\Gamma(\ka\xx, \kb\xx)
&=&
\int dt \,
O_\Gamma^{\rm conf}(\kappa = \kappa_+ + \kappa_- t, t; \xx).
\eea
The coefficient functions 
$F_\Gamma^{\rm conf}(x^2, \kappa, t)$ 
in the conformal operator product expansion, which is 
analogous to equ.~(\ref{GLCE}), are given by
\bea
F_\Gamma (x^2, \xx p_1, \xx p_2)
=
\int d\rho \int dt 
e^{ i\kappa(\xx p_+) }
\delta \Big(t - \frac{\xx p_-}{\xx p_+}\Big)
F_\Gamma^{\rm conf}(x^2, \kappa, t)
\eea
Let us remark that 
\begin{align}
%e^{ i\kappa(\xx p_+)} \delta (t - 
%\hbox{$\frac{\xx p_-}{\xx p_+}$})
%=%&=
e^{ i\kappa(\xx p_+)} |{\xx p_-}|
\delta \big({\xx p_-}t - {\xx p_+}\big)
%\nonumber\\
=:
\pd_{(i\kappa)}\!\int\! \frac{d\rho}{2\pi}
e^{i\kappa(\xx p_+) + i \rho\left[(\xx p_+)t - (\xx p_-)\right] }
\nonumber
\end{align}
defines $\pd_{(i\kappa)}$ (for a precise mathematical definition
see \cite{BGHR87}).

The corresponding local operators are determined according to
\bea
\label{conloc1}
\hspace{-.8cm}
 O_{\Gamma,\, k\ell}^{\rm conf}
&=&
i^\ell \pd^\ell/\pd\kappa^\ell 
\int dt \, C_k^\nu(t) \, 
O_\Gamma^{\rm conf}(\kappa, t; \xx)\Big|_{\kappa=0}\\
\hspace{-.8cm}
\label{conlocal}
&=&
{\bar\psi}(x_1)\Gamma \left(i\xx\pd_+\right)^\ell
C^\nu_k\!\left(\hbox{\large$
\frac{\xx D_-}{\xx\pd_+}$}
\right) 
\psi(x_2)\Big|_{x_1=x_2=0}
\quad{\rm with}\quad
\ell \leq k,
\\
\label{conloc}
\hspace{-.8cm}
O_\Gamma^{\rm conf}(\kappa, t; \xx)
&=&
\sum_{k,\ell =0}^\infty
(-1)^\ell \frac{\kappa^\ell}{\ell!}
N_k^\nu (1-t^2)^{\nu - \frac{1}{2}}
C_k^\nu(t) O\,_{\Gamma,\, k\ell}^{\rm conf}(\ka\xx, \kb\xx),
\eea
where 
\bea
N_k^\nu
=
2^{1-2\nu} 
\frac{\Gamma^2(\hbox{$\frac{1}{2}$})\Gamma(2\nu+k)}
       {\Gamma^2(\nu) (\nu+k) k!}
\nonumber
\eea
is the normalization factor which appears in the completeness
and orthogonality relation of the Gegenbauer polynomials
$C_k^\nu(t)$ of order $k$, index
$\nu = d + s - \frac{1}{2}$ with $d$ and $s$ being
the (canonical) dimension and the spin of the fields $\psi$,
respectively:
\bea
\sum_{k=0}^\infty 
\frac{ (1-t^2)^{\nu - \frac{1}{2}} }{N_k^\nu}
C_k^\nu(t)C_k^\nu(t')
=%\!\!&=&\!\!
\delta(t-t'),
\quad
%\nonumber
%\\
\int_{-1}^{+1}\!\! dt 
(1-t^2)^{\nu - \frac{1}{2}}C_k^\nu(t)C_{k'}^\nu(t)
=%\!\!&=&\!\!
N_k^\nu \delta_{kk'}.
\nonumber
\eea
Let us remark that in order to verify the relation
(\ref{conloc})
between the local and the nonlocal conformal 
%as well as with the usual 
operators the restriction $\ell \geq k$ must be 
annuled and the resulting function has to be analytically
continued with respect to the argument $t$ which, 
from the orthogonality relation, is restricted 
to the intervall $|t| \leq 1$. 

The relations (\ref{conform}) and (\ref{formcon}) between the 
(usual) nonlocal operators and the conformal ones
do not suffer from this drewback.
Therefore,  it is obvious
that the nonlocal conformal light--cone operators
of definite twist are obtained from the original scalar,
vector and tensor operators of definite twist, 
which have been determined in Chapter 3, by
simply inserting them into eq.~(\ref{conform}).
This may be read off also from eq.~(\ref{conlocal})
if one observes that 
\bea
\label{cloc}
\left(i\xx\pd_+\right)^\ell
C^\nu_k\!\left(\hbox{\large$\frac{\xx D_-}{\xx\pd_+}$}\right) 
=
\left(i\xx\pd_+\right)^{\ell-k}
\sum_{r=0}^k 
e_r^k \left(i\xx\pd_+\right)^r (\xx D_-)^{k-r},
\eea
where $e_r^k$ are the expansion coefficients of the 
Gegenbauer polynomials, 
is a sum of terms containing the same number of 
differential operators which are multiplied by
$\xx^{\mu_1}\ldots \xx^{\mu_{\ell}}$. Again, the
fact will be used that the decomposition of a Lorentz
tensor into irreducible ones does not depend on how the
"internal" symmetry of that tensor is build up. The
twist decompositions of each individual term in the
above expansion (\ref{cloc}) will be resummed to get the twist
decomposition of the local conformal operators. From this
the twist decomposition of the nonlocal conformal
operators obtains.

The general result will be
\bea
O_\Gamma^{\rm conf}(\kappa, t; \xx)
=
O_\Gamma^{\rm conf,~tw2}(\kappa, t; \xx)
+
O_\Gamma^{\rm conf,~tw3}(\kappa, t; \xx)
+
O_\Gamma^{\rm conf,~tw4}(\kappa, t; \xx)
\eea
with
\bea
O_\Gamma^{\rm conf,~\tau}(\kappa, t; \xx)
=
\pd_{(i\kappa)}\int \!\!\frac{d\rho}{2\pi}
O_\Gamma^\tau (\kappa - (1-t)\rho, \kappa + (1+t)\rho)
\eea
where $O_\Gamma^\tau(\ka, \kb)$ has to be taken from
the expressions determined in Subsections \ref{scalar}
-- \ref{tensor}, namely 
Eqs.~(3.12) in the scalar case, (3.25), (3.26), (3.41) 
and (3.46) in the vector case, as well as (3.68), (3.69), 
(3.88) and (3.89) in the tensor case.
Here, it should be remarked that for the generalization
to the conformal operators it was essential that the
twist decomposition holds for arbitrary values of 
$\kappa_i$ and not only for special ones like
$(\ka, \kb) = %(0, \kappa)$ or $
(-\kappa, +\kappa)$.

%%%%%%%%%%%%%%%%%%%%%%%%%%%%%%
%% Datei: lazconcl.tex    XXXX
%%%%%%%%%%%%%%%%%%%%%%%%%%%%%%

\section{Conclusions and Outlook}
\renewcommand{\theequation}{\thesection.\arabic{equation}}
\setcounter{equation}{0}

In this paper we introduced an algorithm to decompose
bilocal light--ray operators into operators of well-defined
geometric twist and applied it to the bilocal quark operators
which appear, e.g.,~in virtual Compton scattering in the
generalized Bjorken region. To achieve this we made a
Taylor expansion with respect to both arbitrary space--time
arguments of the nonlocal operators to get the related
local tensor operators. Group theoretically the method 
is based on the (complete) reduction of the corresponding 
tensor representations of the orthrochronous Lorentz group 
into irreducible representations. Then we resummed the
local operators of definite twist to nonlocal ones.
After projection onto the light--cone the (finite) twist 
decomposition of scalar, vector and antisymmetric tensor 
operators results. Thereby we were able to determine 
the trace terms of the nonlocal operators explicitly. 
Because of the fact that the "internal" Lorentz indices 
of the local operators have to be contracted by the 
(symmetric) product of the coordinates $x$ the 
nonlocal operators with definit twist are harmonic
tensor functions. The polynomial basis of these
tensor functions has been determined in Appendix \ref{trace}
for the scalar, vector and (second rank) antisymmetric tensor case. 
If necessary, this could be extended also to symmetric
tensors -- which are of interest if gluon operators 
have to be considered -- and to tensors of higher rank. 

There are many reasons why these operators had to be 
determined. First of all, the more principal one, to make the
twist decomposition rigorous and explicit. This is required
if the renormalization of the light--ray operators should
be considered on a safe basis. Up to now the anomalous 
dimensions of the light--cone operators have been determined 
without checking it explicitly also for the trace terms.
As long as only leading orders are of interest this may
be acceptable. However, if higher perturbational orders 
come into the play one gets serious troubles in 
determining the anomalous dimensions of, say, twist--4
operators because trace terms contribute essentially.

Another point of principal interest, but also with 
practical consequences, obtains if, as partly has been 
done in Subsection 4.1, matrix elements of
light--cone operators with definite twist are
considered.  Whereas up to now the amplitudes 
related to the twist--2 vector operator had to be projected 
onto the scalar case by an additional contraction with $x$ it
is now possible to extract more detailed information
(see \cite{BGR98}). Namely, having determined 
the explicit twist decomposition it is possible, 
at least in principle, to relate the distribution
amplitudes for vector and tensor operators.
Furthermore, additional informations may be obtained
also for operators of higher twist 
and different relationships between them might be derived. 
This, of course, has to be considered in more detail.

A further problem, which however appears to be much easier
to solve, 
is to show how the geometrical twist decomposition which
has been given in this paper relates to the dynamical twist
decomposition of parton distributions given, e.g.,~by
Jaffe and Ji \cite{Jaf92}; see also \cite{Jaf96}. 
Of course, it is to be expected that both decompositions 
coincide in the leading terms, but differ at higher orders.
This difference between `geometric' and `dynamic' twist
decomposition has been discussed in \cite{BBKT98} using the
framework of Ref.~\cite{BB88}; in fact, the consequences
of this difference for twist--3 (in the dynamical sense)
distribution amplitudes of vector mesons are demonstrated
in full detail.

As we have seen, a simple reduction to the scalar case 
is impossible for the twist--3 operators. This will be
 of interest for the structure function $g_2$ in 
polarized deep inelastic 
scattering. The corresponding twist--3 tensor operators
transforming according the the same symmetry class (ii)  
 are given by \cite{Gey96b}\footnote{
In this Reference the trace terms had been ignored; in addition
the equation of motion, $(i\gamma D - m )\psi = 0$,
 has been used to relate the various
possible twist--3 contributions which mix under renormalization}:
\bea
O^{\rm tw 3}_\mu(\ka x, \kb x)
\!\!&=&\!\!
i (\delta^\nu_\mu (x\pd) - x^\nu\pd_\mu)
\int_0^1\!\!du \, %\Big(
\bar\psi(\ka u x) \gamma_\nu \gamma_5
%U(\ka u x, \kb u x) 
\psi(\kb u x),
%- {\rm trace~terms} \Big) 
\nonumber \\
M^{\rm tw 3}_\mu(\ka x, \kb x)
\!\!&=&\!\!
m (\delta^\nu_\mu (x\pd) - x^\nu\pd_\mu)
\int_0^1\!\!du \, u \,%\Big(
\bar\psi(\ka u x) x^\rho\sigma_{\rho\nu} \gamma_5
%U(\ka u x, \kb u x) 
\psi(\kb u x),
%- {\rm trace~terms} \Big) 
\nonumber\\
S^+_\mu(\ka x, \kb x)
\!\!&=&\!\!
\int_0^1\!\!du \, u \,%\Big(
\bar\psi(\ka u x) %U(\ka u x, \tau x)
(x\gamma) x^\nu \!\left[ {\tilde F}_{\mu\nu}(\tau_u x) 
+ i \gamma_5 F_{\mu\nu}(\tau_u x)\right]\!
%U(\tau x, \kb u x) 
\psi(\kb u x),
%- {\rm trace~terms} \Big) 
\nonumber \\
S^-_\mu(\ka x, \kb x)
\!\!&=&\!\!
\int_0^1\!\!du \, {\bar u} \,%\Big(
\bar\psi(\ka u x) %U(\ka u x, \tau x)
(x\gamma) x^\nu \!\left[ {\tilde F}_{\mu\nu}(\tau_u x) 
- i \gamma_5 F_{\mu\nu}(\tau_u x)\right]\!
%U(\tau x, \kb u x) 
\psi(\kb u x),
%- {\rm trace~terms} \Big)
\nonumber 
\eea
with $\tau_u = {\bar u} \ka + u \kb$; for notational
simplicity we omitted the corresponding phase factors
and the trace terms. Here, the Shuryak--Vainshtein 
operators $S^\pm(\ka, \tau_u, \kb)$ in the integrand
introduce additional complications because they
are trilinear operators (being related to $O_\mu$ and 
$M_\mu$ by the equation of motion).
As might be expected they are related to the symmetry
class (ii) but with the following booking of indices
\bea
\bar\psi(0) \gamma_{\{\mu_1}D_{\mu_2}
\ldots D_{\mu_{\ell-1}}D_{[\beta}D_{\mu_\ell ]}
D_{\mu_\ell+1} \ldots D_{\mu_n\}}\psi(0).
\nonumber
\eea
Therefore, these operators can be considered in the
same manner as has been done for the vector (and
tensor) operators of symmetry type (ii).
A detailed treatment of this more evolved
situation remained open.
\\
\medskip
\medskip

\noindent
{\bf\large Acknowledgement}

\noindent
The authors would like to thank Johannes Bl\"umlein
for many useful discussions. One of the authors (B.G.) grateful
acknowledges the kind hospitality during a stay at
Institute of Theoretical Physics, Karl-Franzens-University 
Graz, and another one (D.R.) thankful acknowledges a
stay as guest of Graduate College "Quantum Field Theory"
at Center for Higher Studies, Leipzig University. 
M.L. acknowledges the fine conditions at
Max-Planck-Institute of Metal Research, Stuttgart, 
extended to him during the final stage of this work.
Finally, we thank S. Serowy for a useful hint
on the form of the LC--operators in axial gauge.

%%%%%%%%%%%%%%%%%%%%%%%%
% Datei: lazapndx.tex  %
%%%%%%%%%%%%%%%%%%%%%%%%

\begin{appendix}

\section{Symmetry classes and tensor representations 
%of classical matrix groups
}
\renewcommand{\theequation}{\thesection.\arabic{equation}}
\setcounter{equation}{0}
\label{young}

In this Appendix for the readers convenience
we collect some facts about tensor representations
of the classical matrix groups $G$ and their relation to the
symmetric group $S_n$ which are of relevance in our 
consideration (see, e.g.,~\cite{BR,Boe55, Ham62,PT}). 
Given any irreducible matrix representation, 
\begin{eqnarray}
G \ni g \mapsto {\hat g} \in {\cal B}(V): 
\qquad {\hat g} e_i = e_j\, {a^j}_i,
\end{eqnarray}
on some vector space $V$ (of dimension ${\rm dim}\,V$)
with base $\{e_i,\, i = 1,2,\ldots,{\rm dim}\,V \}$. 
The direct products
of these representations act on the {\em tensor space}
$T^n\, V$ which is build as tensor product of $n$ 
copies of the vector space $V$:
\begin{eqnarray}
T^n\,V
&=&
\left(V \otimes\ldots\otimes V\right).
\nonumber
\end{eqnarray}
The elements of the tensor space $T^n\,V$ are the $n$--times
contravariant %and $m$--times covariant 
{\it tensors}
${\bf t} = t^{i_1\ldots i_n} e_{i_1\ldots i_n}$, 
where $t^{i_1\ldots i_n}$ 
are the {\em components} of ${\bf t}$ in the base 
${e_{i_1\ldots i_n}}
\equiv
\big(e_{i_1}\otimes\ldots\otimes e_{i_n}\big),\;
i_k \in {\[1,\ldots , {\rm dim\,}V \]}
$.~\footnote{
Here, neither covariant tensors which 
act on direct products of the dual space $V^*$ nor mixed
ones are considered since, for the case under consideration, 
the indices may be raised or lowered by the metric tensor.
}
The components of the tensors of rank $n$ transform under
$G$ according to
\begin{eqnarray}
\label{tensortrf}
(t')^{i_1\ldots i_n}
\equiv
t^{i'_1\ldots i'_n}
=
{a^{i'_1}}_{i_1}
\ldots 
{a^{i'_n}}_{i_n}
t^{i_1\ldots i_n}.
\end{eqnarray}

In the following we shall comment on the connection between the
irreducible representations of the general linear group 
$GL(N, {\Bbb C})$ (together with
its subgroups) and of the symmetric group $S_n$ on 
these tensor spaces.
Let us, at the moment consider only the groups 
$G = GL(N,{\Bbb C}),SL(N,{\Bbb C}), SU(N)$.
Given any (irreducible) representation of $G$ on  
$V$, then its $n$--fold tensor product defines a reducible
representation $A^{(n)}$ which, together with the 
reducible representation $\hat \pi$ of $S_n$, is 
determined by:
\begin{eqnarray}
{\hat A}^{(n)}(g){\bf t}
\equiv \;
{\hat A}^{(n)}(g)(t^{i_1\ldots i_n}e_{i_1\ldots i_n})
&:=&
t^{i_1\ldots i_n}
({\hat g}e)_{i_1}\otimes\ldots\otimes ({\hat g}e)_{i_n},
\quad\forall g \in G,
\nonumber
\\
{\hat \pi}{\bf t}
\equiv 
~~\qquad
{\hat \pi}(t^{i_1\ldots i_n}e_{i_1\ldots i_n})
&:=&
t^{\pi(i_1\ldots i_n)}
e_{i_1}\otimes\ldots\otimes e_{i_n},
\quad \forall \pi \in S_n;
\nonumber
\end{eqnarray}
therefore the action of both groups on the tensors 
$\bf t$ commutes~\footnote{
This results from the fact that the product of the 
matrix elements $a^j_i$ in eq.~(\ref{tensortrf})
is {\em bisymmetric} under $S_n$.
}
\begin{eqnarray}
\label{Schur}
\big[ {\hat A}^{(n)}(g), {\hat \pi} \big] {\bf t} = 0.
\end{eqnarray}
Of course, the same conclusion holds for the elements 
$\sum_\pi \alpha(\pi) {\hat \pi}, \alpha(\pi) \in {\Bbb R}$, 
of the group algebra ${\cal R}[S_n]$ of the symmetric group. 
This algebra, considered as a vector space, carries the 
{\em regular representation} of $S_n$ which is known to be
fully reducible. 

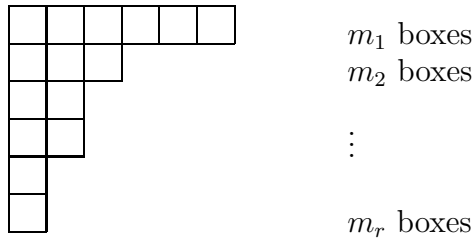
\begin{figure}[h]
\unitlength0.5cm
\begin{center}
\begin{picture}(15,5)
\linethickness{0.15mm}
\multiput(3,4)(1,0){5}{\line(0,1){1}}
\multiput(1,-1)(1,0){2}{\line(0,1){6}}
\multiput(3,1)(3,0){1}{\line(0,1){4}}
\multiput(4,3)(4,0){1}{\line(0,1){2}}
\put(1,5){\line(1,0){6}}
\put(1,4){\line(1,0){6}}
\put(1,3){\line(1,0){3}}
\put(1,2){\line(1,0){2}}
\put(1,1){\line(1,0){2}}
\put(1,0){\line(1,0){1}}
\put(1,-1){\line(1,0){1}}
\put(10,4){$m_1$ boxes}
\put(10,3){$m_2$ boxes}
\put(10,1){$\vdots$ }
\put(10,-1){$m_r$ boxes}
\end{picture}
\end{center}
\caption{\label{Y-Rahmen} Young pattern $\underline m$} 
\end{figure}

The irreducible representations $\Delta^{[m]}$ of the 
symmetric group are uniquely determined by the idempotent 
(normalized) Young operators
\begin{eqnarray}
\label{Young1}
{\cal Y}_{[m]} &=& \frac{f_{[m]}}{n!}{\cal P}{\cal Q}
\quad {\rm with} \quad
{\cal P} =\!\!\! \sum_{p \in H_{[m]}} p,
\quad 
{\cal Q} ~= \sum_{q \in V_{[m]}} \delta_q\,q,
\\
\label{Young2}
{\cal Y}_{[m]}{\cal Y}_{[m']}
&=&
\delta_{[m][m']}{\cal Y}_{[m]},
%\nonumber
\end{eqnarray}
which are related to corresponding Young tableaux
being  denoted by $[m]$.
If
\begin{eqnarray}
{\underline m} = (m_1, m_2, \ldots m_r) 
\quad
{\rm with}
\quad
m_1 \geq m_2\geq \ldots \geq m_r,
\quad
%{\rm and}
%\quad
\sum^r_{i=1}\, m_i = n,
\end{eqnarray}
defines a Young pattern (Fig.~\ref{Y-Rahmen}) then a {\em 
Young tableaux} $[m]$ is obtained by putting in (without
repetition) 
the indices $i_1, \ldots i_n$ -- corresponding to different
``places''
within the direct product -- and $H_{[m]}$ and $V_{[m]}$ denotes 
their horizontal and vertical permutations with respect to
$[m]$.
A {\em standard tableaux} is obtained when the indices
$i_1, \ldots, i_n$ are ordered lexicographically. There are
\begin{eqnarray}
\label{f}
f_{[m]} = n! \, 
\frac{\prod_{i<j} (l_i - l_j)}{\prod_{i=1}^r l_i!}
\quad
{\rm with}
\quad
l_i = m_i + r - i,
\quad
%{\rm and}
%\quad
\sum_{[m]}f_{[m]}^2 = n!,
\end{eqnarray}
different standard tableaux which correspond to $f_{[m]}$ 
different, but equivalent, irreducible representations of $S_n$
whose dimension is given also by $f_{[m]}$.
The (normalized) Young operators ${\cal Y}_{[m]}$ 
according to (\ref{Young2}) project onto 
mutually orthogonal irreducible left ideals of ${\cal R}[S_n]$.
%%%%%%%%%%%%%%%%%%%%%%%%%%%%%%%%%%%%%%%%%%%%
%%%%%%%%%%%%%%%%%%%
%%%%%%%%%%%%%%%%%%%%%%%%%%%%%%%%%%%%%%%

In this way the tensor product of $n$ spaces $V$ carries the 
regular representation of the group $S_n$ and decomposes into 
$\sum_{[m]} f_{[m]}$ irreducible subspaces containing tensors 
of symmetry class $[m]$. However, the tensors actually are
specified by taking for the indices $i_1, \ldots, i_n$ 
arbitrary values from the range $[1, \ldots, {\rm dim}\,V]$.
Therefore, with respect to the group $G$ the standard tableaux 
is defined by putting into the Young pattern the {\em values of
the indices} $i_k$ such that they are non--decreasing from
left to right and increasing from top to bottom!
Because of eq.~(\ref{Schur}) by Schur's Lemma it follows that
the representation ${\hat A}^{(n)}(g)$ of $G$ is reducible and 
it decomposes into as many irreducible representations as there
are (in general reducible) representations of $S_n$, i.e., any
tensor
representation of the group $G$ is characterized by a
standard tableaux,
\begin{eqnarray}
{\hat A}^{(n)}(g) 
=
\underset{[m]}{\oplus}
R^{[m]}(g) \otimes E_{l_{[m]}},
\quad
{\hat \pi}
=
\underset{[m]}{\oplus}
 E_{r_{[m]}} \otimes \Delta^{[m]}(\pi),
\end{eqnarray}
where $E_{l_{[m]}}$ is the unit matrix; %in the left ideal 
%carrying the irreducible representation of $S_n$
 more explicitly we have
\begin{eqnarray}
\label{tenrep}
{\hat A}^{(n)}(g)_{[m] \kappa \alpha, [m'] \kappa' \alpha'}
&=&
\delta_{[m][m']} 
\delta_{\alpha \alpha'}
R^{[m]}_{\kappa \kappa'}(g),
\end{eqnarray}
where $\kappa = 1, \ldots , r_{[m]}$ counts different
equivalent irreducible representations of $S_n$ and 
$ \alpha = 1, \ldots , l_{[m]} \leq f_{[m]}$ counts the 
basis elements of these representations. The  
representations $R^{[m]}(g)$ of $G$ in $T^n\,V$ are 
characterized by equivalent Young operators 
${\widehat{\cal Y}}_{[m]} = {\cal Q}{\cal P}$:~\footnote{
Note, that the symmetrizations $\cal P$ and the
antisymmetrizations $\cal Q$ with respect to $[m]$
are interchanged. This corresponds, in the terminology
of quantum mechanics, to ``quantum number permutations''
instead of the ``place permutations'' above. And, contrary
to ${\cal Y}_{[m]}$ which defines a left ideal in $\cal R$,
${\widehat{\cal Y}}_{[m]}$ defines a right ideal.
}
\begin{equation}
\label{dsum}
T^n\,V
~=~
\underset{[m]}{\oplus}
\underset{\kappa}{\oplus}
{\widehat{\cal Y}}_{[m]\kappa}(T^n\,V).
\end{equation}
%Of course, every irreducible representation $R^{[m]}(g)$
%of $G$ 
%has dimension $f_{[m]}$ and is contained $f_{[m]}$--times
%in the (reducible) representation eq.~(\ref{tenrep}).
The invariant subspaces projected out by
${\widehat{\cal Y}}_{[m]\kappa}(T^n\,V)$ are irreducible with 
respect to $GL(N,{\Bbb C}), SL(N,{\Bbb C})$ and $SU(N)$. 
After restriction onto these subgroups some of 
the irreducible representations become equivalent ones. 
 However, for $O(N,{\Bbb C}), SO(N,{\Bbb C})$ and
$Sp(2\nu)$ these representations, in general, are not
irreducible and decompose further.  

Concerning the orthogonal groups from the subspaces of
the above introduced symmetry classes only the 
completely antisymmetric ones (for $n \leq N$) remain 
irreducible. The reason is that, because of the very definition 
of the orthogonal groups,
$\delta_{ij} {a^i}_k {a^j}_l 
~=~ \delta_{kl}, \;\forall a \in O(N)$, 
the operation of taking the trace of a tensor 
commutes with the orthogonal transformations of that tensor:
\begin{eqnarray}
\label{decomp}
{\rm tr}\; {\bf T}'
~=~ \delta_{ij} {T'}^{ij}
~=~ \delta_{ij} {a^i}_k {a^j}_l T^{kl}
~=~ {\rm tr}\; {\bf T}.
\end{eqnarray}
Again, by Schur's Lemma, irreducible subspaces 
of $O(N)$ are spanned by {\em traceless}
tensors having definite symmetry class.
This decomposition is obtained as follows:
\begin{eqnarray}
\label{TL}
T^{i_1 i_2 \ldots i_n}_{[m]}
&=&
\TL {T_{[m]}^{i_1 i_2 \ldots i_n}}
+ \sum\limits_{1 \leq r,s \leq n} \delta^{i_r i_s}\, 
T^{i_1 \ldots i_{r-1}i_{r+1}\ldots 
i_{s-1}i_{s+1}\ldots i_n}_{[m-2]}.
%\nonumber
\end{eqnarray}
The tensors which appear under the sum have degree $n-2$ and a
 Young pattern ${[m-2]}\subset {[m]}$ obtained by removing 
two boxes from the (right) border without destroying the property 
(\ref{even}) and (\ref{odd}) to be a pattern. They
 may be decomposed again into traceless ones plus some
remainder, and so on. Therefore, a traceless tensor is obtained
from the original one by succesively subtracting the traces.

%\pagebreak

\section{Harmonic tensor functions }
\renewcommand{\theequation}{\thesection.\arabic{equation}}
\setcounter{equation}{0}
\label{trace}

In this Appendix we derive the projection operators which 
determine the traceless part of
 a completely symmetric tensor of rank $n$, 
$T_{\{\mu_1 \ldots \mu_n\}}$, of a tensor of rank
$n+1$ being symmetric in $n$ of its indices,
$T_{\alpha\{\mu_1 \ldots \mu_n\}}$, and of a
tensor of rank $n+2$ being symmetric in $n$ and antisymmetric
in the remaining two indices,
$T_{[\alpha\beta]\{\mu_1 \ldots \mu_n\}}$. To achive
this we use the fact that, after contracting the indices 
of the symmetric part with some vector $x$, 
the resulting scalar, vector and antisymmetric tensor functions
obey the following equations:
\begin{eqnarray}
\label{H0}
\square \tl T_n(x) \!\!\!&=&\!\!\! 0,
\\
\label{H1}
\square \tl T_{\alpha n}(x) \!\!\!&=&\!\!\! 0,
\qquad~~\,
\pd^\alpha \tl T_{\alpha n}(x) = 0,
\\
\label{H2}
\square \tl T_{[\alpha \beta] n}(x) \!\!\!&=&\!\!\! 0,
\qquad
\pd^\alpha \tl T_{[\alpha \beta] n}(x) = 0,
\qquad
\pd^\beta  \tl T_{[\alpha \beta] n}(x) = 0.
\end{eqnarray}

\noindent
(i)\hspace{.5cm}
The first of these equations define the {\em harmonic polynomials} of
order $n$. Its solution, in $d$ dimensions, is given by
(see e.g. \cite{Vil}, Chapter IX)
\begin{eqnarray}
\label{T_harm_d}
\tl T_n(x) 
&=&
H^{(d)}_n\!\left(x^2|\square\right)\!T_n(x)\\
&=&
\bigg\{1 +
\sum_{k=1}^{[\frac{n}{2}]}
\bigg(\prod_{\ell=1}^k \frac{1}{d+2n-2\ell-2}\bigg)
\frac{(-x^2)^k\,\square^{k}}{2^k\,k!}\bigg\} T_n(x),
\nonumber
\end{eqnarray}
which, for $d=4$, may be rewritten as
\begin{eqnarray}
\label{T_harm4}
\tl T_n(x) 
=
H^{(4)}_n\!\left(x^2|\square\right)T_n(x)
\end{eqnarray}
with the harmonic projection operator
\begin{eqnarray}
\label{Harm4}
H^{(4)}_n\!\left(x^2|\square\right)
=
\sum_{k=0}^{[\frac{n}{2}]}
\frac{(n-k)!}{k!n!}
\left(\frac{-x^2}{4}\right)^{\!\!k}
\square^{k} .
%\quad
%\mathrm{with}
%\quad
%\square \, H^{(4)}_n\!\left(x^2|\square\right) \equiv 0.
\end{eqnarray}

\noindent
(ii)\hspace{.5cm}
Let us now consider the set of equations (\ref{H1}) to determine
the harmonic vector functions $T_{\alpha n}(x)$ being
related to the traceless tensors 
$T_{\alpha\{\mu_1 \ldots \mu_n\}}$ which, to the best of our
knowledge, are not determined up to now.
Of course, after the $n$--fold contraction with $x$ they
are harmonic polynomials of order $n$, 
\begin{eqnarray}
T_{\alpha \tl n}(x)
:= 
H^{(4)}_n\!\left(x^2|\square\right)
T_{\alpha n}(x), 
\end{eqnarray}
and the traces corresponding to $g_{\alpha\mu_i}$
are obtained through differentiation with respect to
$x_\alpha$. Therefore, the most general ansatz for the
harmonic vector functions which conserves its order is 
\begin{eqnarray}
\tl T_{\alpha n}(x)
= 
\left\{\delta_{\alpha}^{\beta}
-a_n x_\alpha \pd^{\beta}(x\pd)
   -   b_n x^2\pd_\alpha\pd^{\beta}
\right\}
H^{(4)}_n\!\left(x^2|\square\right)
T_{\beta n}(x). 
\end{eqnarray}
From the two conditions (\ref{H1}) we obtain two equations
from which the coefficients $a_n,\ b_n$ can be determined:
\begin{eqnarray}
1 &=& n (d + n - 1) a_n + 2(n - 1) b_n,
\nonumber
\\
0 &=& n a_n + (d + 2n - 4) b_n.
\nonumber
\end{eqnarray}
Note, that to derive these conditions we used the fact that
$\square H^{(4)}_n\!\left(x^2|\square\right) \equiv 0$,
$\big[(x\pd),H^{(4)}_n\big] = 0$ and 
$(x\pd) T_{\beta n}(x) = n T_{\beta n}(x)$.
%\pagebreak
The solution of these equations is
\begin{eqnarray}
n\ a_n = \frac{d+2n-4}{(d+2n-2)(d+n-3)},
\qquad
b_n = - \frac{1}{(d+2n-2)(d+n-3)};
\nonumber
\end{eqnarray}
for $d=4$ it reduces to
\begin{eqnarray}
a_n = 1/(n+1)^2,
\qquad
b_n = - a_n/2.
\nonumber
\end{eqnarray}
 Therefore,
we obtain the following result:
\begin{equation}
\label{Proj}
\tl T_{\alpha n}(x)=\left\{\delta_{\alpha}^{\beta}
-\frac{1}{(n+1)^2}
\left[x_\alpha\pd^\beta(x\pd)
-\hbox{\large$\frac{1}{2}$} x^2\pd_\alpha\pd^\beta\right]
\right\}
H^{(4)}_n\!\left(x^2|\square\right) T_{\beta n}(x)\, .
\end{equation}
If eq.~(\ref{Proj}) will be contracted with $x^\alpha$ 
then, after some tedious but straightforward calculations,
we obtain, as is should be, the harmonic polynomial 
$\tl T_{n+1}(x)$.
%\begin{eqnarray}
%\label{Proj2}
%\tl T_{n+1}(x) &=&
%\sum_{k=0}^{[\frac{n+1}{2}]}
%\frac{(n+1-k)!}{k!(n+1)!}\left(
%\frac{-x^2}{4}\right)^k
%\square^k T_{n+1}(x)\ .
%\nonumber
%\end{eqnarray}
%Let us remark that this expression may written with the help
%of the generators $P_\beta$ and $K_\alpha$ of the conformal
%group (for a -- spinless! -- operator of dimension $d=1$):
%\begin{equation}
%\label{Proj3}
%\tl T_{\alpha n}(x)
%=\left\{\delta_{\alpha}^{\beta}
%+\hbox{\large$\frac{1}{2(n+1)^2}$}
%K_\alpha P^\beta\right\}
%H^{(4)}\!\left(x^2|\square\right)T_{\beta n}(x)\, .
%\end{equation}
%

\medskip
\noindent
(iii)\hspace{.5cm}
Using the same procedure we may construct the harmonic
antisymmetric tensor functions $\tl T_{[\alpha\beta] n}(x)$.
The most general ansatz for it has the following structure
(remind the convention $T_{[\mu\nu]} = \frac{1}{2}
(T_{\mu\nu}-T_{\nu\mu})$):
\begin{equation}
\hspace{-.1cm}
\tl T_{[\alpha\beta] n}
=
\left\{\delta_\alpha^\mu\delta_\beta^\nu
-a_n x_{[\alpha}\delta_{\beta]}^\nu\pd^\mu
-b_n x^2\pd^\mu\pd_{[\alpha}\delta_{\beta]}^\nu
-c_n x^\mu x_{[\alpha}\pd_{\beta]}\pd^\nu
\right\}
%H^{(4)}\left(x^2|\square\right) 
T_{[\mu\nu] \tl n}(x).
\end{equation}
The conditions (\ref{H2}) lead to the requirement
that the coefficients of the terms 
$\delta_\beta^{\mu}\pd^\nu, x^\mu\pd^\nu\pd_\beta$ and 
$\pd_{[\alpha}\delta_{\beta]}^\mu\pd^\nu$ should vanish. 
The corresponding system of linear equations reads
\begin{eqnarray}
1&=&(d+n-2)a_n/2 + (n-1)b_n \nonumber\\
0&=&2b_n+(d+n-2)c_n      \nonumber\\
0&=&a_n+c_n+(d+2n-4)b_n\, .   \nonumber
\end{eqnarray}
It is solved by the following values:
\begin{eqnarray}
(a_n + c_n, \;b_n, \;c_n)
=\frac{2}{(d+2n-2)(d+n-4)}
\left(d+2n-4,\; -1,\; \frac{2}{(d+n-2)}\right),
\nonumber
\end{eqnarray}
and for $d=4$ we obtain
\begin{eqnarray}
(a_n + c_n, \;b_n, \;c_n)
=\frac{2}{n(n+1)(n+2)}
\left(n(n+2),\;-\hbox{\large$\frac{1}{2}$}(n+2),\;1\right)\ .
\nonumber
\end{eqnarray}
From this, using the replacement
$x^\mu x_{[\alpha}\pd_{\beta]}\pd^\nu
= x_{[\alpha}\pd_{\beta]} x^\mu \pd^\nu
- x_{[\alpha}\delta_{\beta]}^\mu\pd^\nu$
 we finally obtain:
\begin{align}
\label{Proj5}
%\lefteqn{
\tl T_{[\alpha\beta] n}(x)
=&
%\hbox{\large$\frac{1}{2}$}
\Big\{\delta_{[\alpha}^\mu\delta_{\beta]}^\nu
-\hbox{\large$\frac{2}{n(n+1)(n+2)}$}
\Big( x_{[\alpha}\pd_{\beta]}x^{[\mu}\pd^{\nu]}%}
\\
& -
 \left[ x_{[\alpha}\delta_{\beta]}^{[\mu}\pd^{\nu]}(x\pd)
-\hbox{\large$\frac{1}{2}$}
x^2\pd_{[\alpha}\delta_{\beta]}^{[\mu}\pd^{\nu]}\right]
(x\pd+2)\Big)\Big\}
H^{(4)}_n\!\left(x^2|\square\right)T_{[\mu\nu]n}(x)\, .%\qquad
\nonumber
\end{align}
%
%This equation may be rewritten by the help of the 
%conformal generators as:
%\begin{align}\label{Proj6}
%\tl T_{[\alpha\beta] n}(x)&=
%\hbox{\large$\frac{1}{2}$}
%\left\{\delta_{[\alpha}^\mu\delta_{\beta]}^\nu
%+\hbox{\large$\frac{1}{n(n+1)(n+2)}$}
%\left(\! M_{\alpha\beta}M^{\mu\nu}
%+\hbox{\large$\frac{1}{2}$}
%K_{[\alpha}P^{[\mu}\delta_{\beta]}^{\nu]}
%\left(1+ D\right)\!\right)\!\right\}\!
%H^{(4)}\left(x^2|\square\right) 
%T_{[\mu\nu] \tl n}(x) .
%\end{align}
Let us remark that the various differential operators which
appear in eqs.~(\ref{Proj}) and (\ref{Proj5}) may be expressed
by the
generators of the conformal group, namely 
$P_\mu, M_{\mu\nu}, D$ and $K_\mu$.
\end{appendix}

%%%%%%%%%%%%%%%%%%%%%%%%%%%%%%
%% Datei: lazrefer.tex   %%%%%
%%%%%%%%%%%%%%%%%%%%%%%%%%%%%%

\end{document}